\documentclass[notoc]{JHEP3}
\usepackage{amsmath}
\usepackage{graphicx}

\title{ Inverse Soft Factors and Grassmannian Residues}
\author{Mathew Bullimore\\ 
			Rudolf Peierls Centre for Theoretical Physics,\\
	 		1 Keble Road, Oxford, OX1 3NP,\\
			United Kingdom\\
			E-mail: \email{m.bullimore1@physics.ox.ac.uk}}

\abstract{The action of inverse soft factors on scattering amplitudes in $\cN=4$ SYM is shown to take a remarkably simple form in momentum twistor space. This is used to identify individual residues of the grassmannian with primitive leading singularities at NMHV and $\Nsq$ and to derive explicit expressions in terms of momentum twistors.}

\def \be  {\begin{equation}}
\def \ee  {\end{equation}}
\def \ba  {\begin{eqnarray}}
\def \ea  {\end{eqnarray}}
\def \bb  {\begin{thebibliography}}
\def \eb  {\end{thebibliography}}
\def \lab #1 {\label{#1}}
\def \ni   {\noindent}
\def \nn  {\nonumber}

\newcommand\cA{\mathcal{A}}
\newcommand\cB{\mathcal{B}}
\newcommand\cC{\mathcal{C}}

\newcommand\cO{\mathcal{O}}

\newcommand{\cR}{\mathcal{R}}

\newcommand\lb{\lambda}
\newcommand\tlb{\tilde{\lambda}}
\newcommand\al{\alpha}
\newcommand\dal{\dot{\alpha}}

\newcommand\Nsq{\mathrm{N}^2\mathrm{MHV}}
\newcommand\NMHV{\mathrm{NMHV}}
\newcommand\MHV{\mathrm{MHV}}
\newcommand\Np{\mathrm{N}^{p}\mathrm{MHV}}

\newcommand\cN{\mathcal{N}}
\newcommand\C {\mathbb{C }}

\newcommand\bP{\mathbb{P}}
\newcommand\CP {\mathbb{CP}}

\newcommand\rd{\mathrm{d}}
\newcommand\rD{\mathrm{D}}

\newcommand\la{\langle}
\newcommand\ra{\rangle}

\newcommand\MHVbar{\overline{\mbox{MHV}}}

\begin{document}



\section{Introduction}

A remarkable formula has been found for scattering amplitudes in $\cN=4$ SYM~\cite{ArkaniHamed:2009dn}. For scattering amplitudes of $n$-particles and degree N$^{k-2}$MHV the formula is a contour integral in the grassmannian $G(k,n)$ whose residues are conjectured to compute all leading singularities of scattering amplitudes in $\cN=4$ SYM. The formula may be written

\begin{equation}
\label{TwistorGrassmannian}
A(Z_1,\ldots, Z_n) = \frac{1}{\mathrm{Vol}\,GL(2)} \int \frac{d^{k\times n}C}{(1)\ldots(n)} \prod_{r=1}^k \delta^{4|4}(C_{ri}Z_i)\,,
\end{equation}

\ni where external states are described by twistors $Z^{\cA}_i=(\tilde{\mu}^{\al}_i,\tlb_{i\dal},\eta^A_i)$ and $C_{ri}$ are homogeneous coordinates on the grassmannian. Individual residues describe primitive leading singularities of loop amplitudes and residue theorems provide all of the important relationships between them~\cite{ArkaniHamed:2009dn}. 

Contours that compute tree-level amplitudes may be constructed in a physical way with a particle interpretation in the grassmannian~\cite{ArkaniHamed:2009dg}. There are many ways to evaluate the tree-level contours that are related by residue theorems. For example, many choices correspond a BCFW expansion of the tree amplitude, where each residue is a primitive leading singularity of a $p$-loop amplitude at N$^p$MHV~\cite{Bullimore:2009cb}. Another residue theorem leads to CSW rules for the tree-level NMHV amplitude and more generally to the Risager expansion for tree amplitudes of higher degree~\cite{ArkaniHamed:2009sx}.

In~\cite{Mason:2009qx}, an alternative grassmannian integral has been proposed that manifests dual superconformal invariance, and where external states are described by momentum twistors~\cite{Hodges:2009hk}. This has been derived from equation~\eqref{TwistorGrassmannian} in~\cite{ArkaniHamed:2009vw}. It has also been shown~\cite{Drummond:2010qh} that the grassmannian formula transforms to a total derivative under the Yangian generators~\cite{Drummond:2009fd},
\begin{equation}
{j^{\cA}}_{\cB} = \sum\limits_{i=1}^n Z_i^{\cA} \frac{\partial}{\partial Z_i^{\cB}}  
\end{equation}
\begin{equation}
{j^{(1)\cA}}_{\cB} = \sum\limits_{i< j}(-1)^{\cC}\left[ Z^{\cA}_i \frac{\partial}{\partial Z_i^{\cC}}  Z^{\cC}_j \frac{\partial}{\partial Z_j^{\cB}}	- (i\leftrightarrow j)\right]
\end{equation}

\ni and that this property uniquely determines the form of the integrand~\cite{Drummond:2010uq,Korchemsky:2010ut}. The individual residues of the grassmannian integral are therefore Yangian invariants. 

The identification of residues of the grassmannian integral with leading singularities is an important problem that is the focus of this paper. We show that inverse soft factors~\cite{ArkaniHamed:2009dn}, which create new leading singularities by adding additional particles, have a very simple action on momentum twistors. We also explain how an inverse soft factor acts on grassmannian residues and use this to identify large classes of residues with primitive leading singularities at NMHV and $\Nsq$. The simple action on momentum twistors then allows explicit expressions for the corresponding Yangian invariants to be written down.


\section{Inverse Soft Factors}
\label{InverseSoftLimits}

In this section we introduce the operation of an inverse soft factor~\cite{ArkaniHamed:2009dn} and show that it has a very simple action on scattering amplitudes when written in terms of momentum twistors~\cite{Hodges:2009hk}.


\subsection{Momentum Space}


We consider scattering amplitudes in $\cN=4$ SYM with external states labelled in on-shell superspace by $i =\{\lb^{\al}_i,\tlb^{\dal}_i,\eta^A_i\}$. The null four-momenta of external particles are then $p^{\al\dal}_i = \lb^{\al}_i\tlb^{\dal}_i$ and fermionic parts of the supermomenta are $q_i^{\al A} = \lb^{\al}_i\eta^A_i$. Here we will consider mainly leading singularities which are rational functions of the kinematic variables. Since tree amplitudes may be expressed as sums of leading singularities then all statements may equally be applied to tree amplitudes~\cite{Britto:2005fq,Britto:2004ap}. Let us denote a generic leading singularity with $n$ particles by
\begin{equation}
		\cO_n(1,\ldots,n).
\end{equation}


An inverse soft factor takes an $n$-particle leading singularity and forms a new leading singularity of the same MHV degree with $(n+1)$ particles. Consider any leading singularity $\cO_n(a,b,\ldots)$ where particles $a$ and $b$ are adjacent, then following~\cite{ArkaniHamed:2009dn} we define the inverse soft limit that adds particle $c$ in between $a$ and $b$ by the following formula,
\begin{equation}
\label{ISLdef}
		\cO_{n+1}(a,c,b,\ldots) \equiv \frac{ \la ab \ra}{ \la ac \ra\la cb \ra} \cO_{n}(a',b',\ldots)
\end{equation}

\ni The primed labels in equation~\eqref{ISLdef} denote shifted external variables $i'=\{\lb_i,\tlb'_i,\eta'_i\}$ where the right-handed spinors and grassmann parameters have been shifted
\begin{eqnarray}
\label{ISLshift1}
		\tlb_a' = \tlb_a + \frac{\la cb\ra}{\la ab \ra}\tlb_c \hspace{1cm} \tlb_b' = \tlb_b+\frac{\la ca\ra}{\la ba \ra}\tlb_c \nn\\
		\eta'_a = \eta_a + \frac{\la cb\ra}{\la ab \ra}\eta_c \hspace{1cm} \eta'_b = \eta_b+\frac{\la ca\ra}{\la ba \ra}\eta_c\, .
\end{eqnarray}

\ni A straightforward application of the Schouten identity confirms that momentum and supermomentum are conserved in the inverse soft factor. 

The inverse soft factor is designed so that

\begin{equation}
\label{softfactor}
	\cO(a,c,b,\ldots) \longrightarrow \frac{ \la ab \ra}{ \la ac \ra\la cb \ra}\cO(a,b,\ldots) \hspace{0.5cm} \mbox{as} \hspace{0.5cm} \lb_c \longrightarrow 0 \, .
\end{equation}

\ni In words, the inverse soft factor creates a leading singularity whose limit as the additional particle becomes soft reproduces the original. The soft behaviour in equation~\eqref{softfactor} comes entirely from the tree-level MHV superamplitude

\begin{eqnarray}
		A^{\MHV}(1,\dots,n) = \frac{1}{\la 12 \ra\ldots\la n1 \ra}\delta^4(\sum_{i=1}^{n}\lb_i\tlb_i) \delta^8(\sum_{i=1}^{n}\lb_i \eta_i)\, ,
\end{eqnarray}

\ni which appears as a pre-factor in all superamplitudes in $\cN=4$ SYM. Therefore the inverse soft factor should add another particle to this superamplitude, changing $A^{\MHV}(1,\ldots,n)$ into $A^{\MHV}(1,\ldots,n+1)$. This follows simply from the conservation of momentum and from the prefactor in the definition~\eqref{ISLdef}. In the following we solve momentum and supermomentum conservation and remove the tree-level MHV superamplitude prefactor. Therefore we consider only the shift~\eqref{ISLshift1} of the kinematic variables.


\subsection{Region Momenta}

We now implement the inverse soft limit in terms of region momenta~\cite{Drummond:2008vq}. The region momenta $\{x_i,\theta_i\}$ are introduced by solving supermomentum conservation
\begin{equation}
\label{dualdef}
		\lb_i \tlb_i = x_{i+1} - x_i \hspace{1cm} \lb_i \eta_i = \theta_{i+1} - \theta_i \, ,
\end{equation}

\ni where $x_{n+1} \equiv x_1$ and $\theta_{n+1}\equiv \theta_1$. The region momenta are then null separated
\begin{equation}
		(x_{i+1}-x_i)^2=0 \hspace{1cm} (\theta_{i+1}-\theta_i)^2=0
\end{equation}

\begin{figure}[h]
\centering
\includegraphics[height=4.5cm]{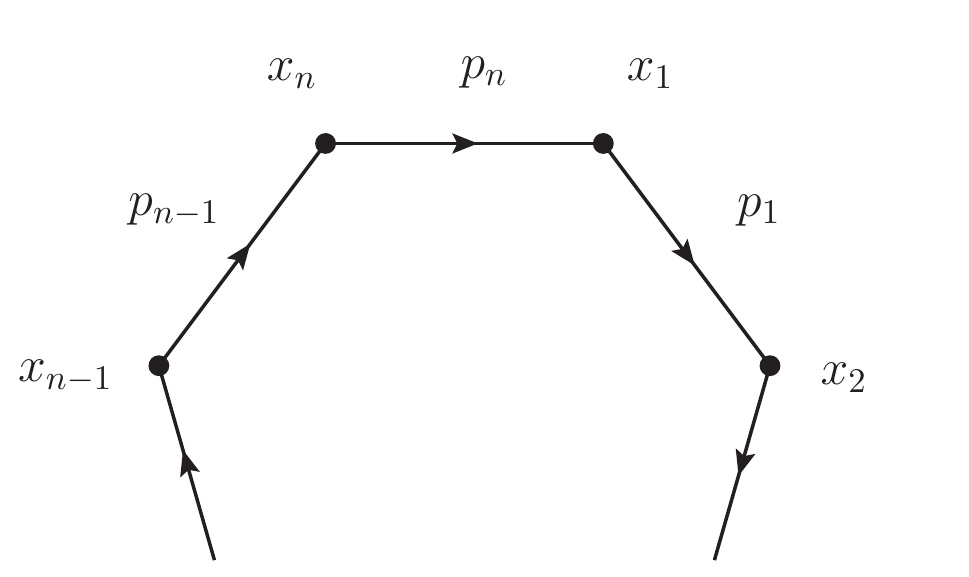}
\caption{\emph{The null polygon formed by the region momenta coordinates.}}
\label{fig:Dualspace}
\end{figure}

\ni and form a null polygon in region momentum space (see figure~\ref{fig:Dualspace}). The primary coordinates are now $\{\lb_i,x_i,\theta_i\}$ and the right-handed spinors $\{\tlb_i\}$ become secondary derived variables.

\begin{figure}[h]
\centering
\includegraphics[height=5.5cm]{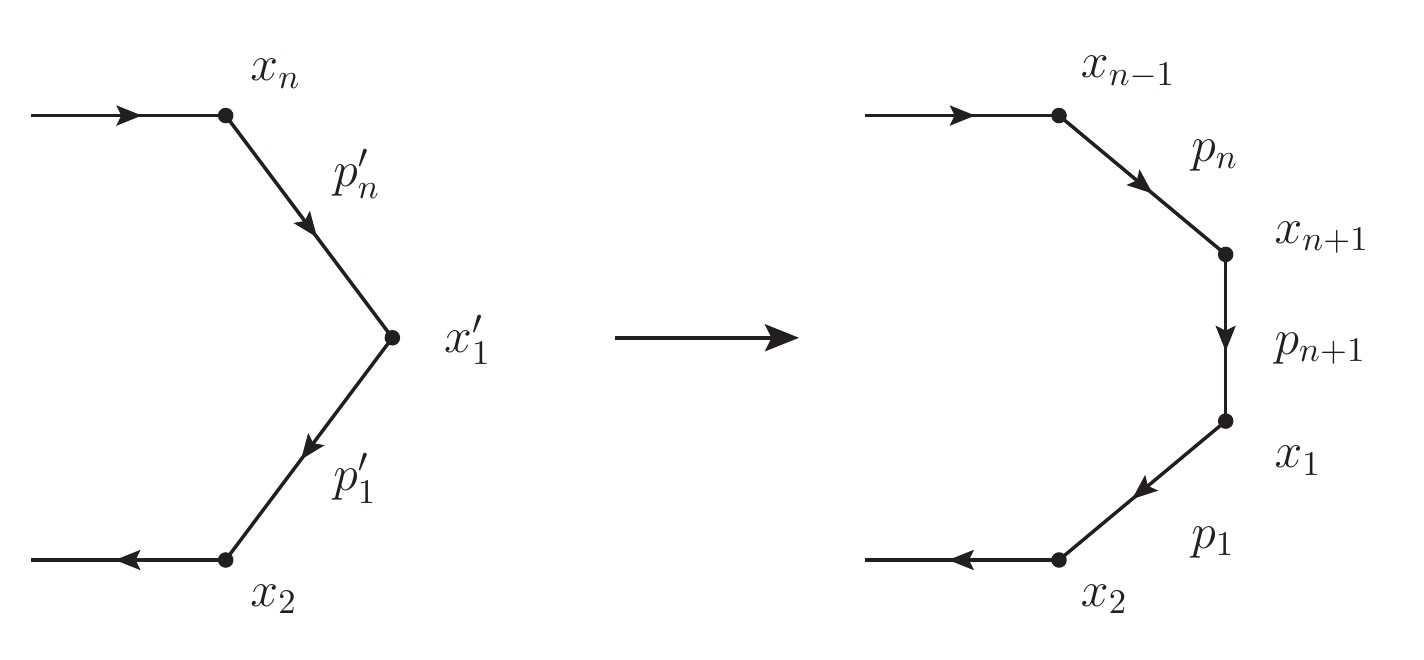}
\caption{\emph{The action of an inverse soft factor on the null polygon in region momentum space.}}
\label{fig:ISLdual}
\end{figure}

Since the region momenta solve supermomentum conservation we remove the MHV prefactor and focus on the remainder, which we denote by $R(\lb_i,x_i,\theta_i)$. Now consider adding the particle $(n+1)$ with an inverse soft factor - see figure~\ref{fig:ISLdual}. The only variables to be shifted are those associated to particles $n$ and $1$. Therefore, to maintain supermomentum conservation, only the region momentum $(x_1,\theta_1)$ may be shifted. Suppressing the unshifted variables we have

\begin{equation}
		 R_{n}(x_1,\theta_1) \longrightarrow R_{n+1}(x_{n+1},\theta_{n+1}; x_1,\theta_1) \equiv R_{n}(x_1',\theta_1')\end{equation}
\vspace{0.1mm}

\ni where the shifted region momenta $\{x_n',\theta_n'\}$ are to be determined. Now referring to figure~\ref{fig:ISLdual}, momentum conservation requires that  
\begin{equation}
\label{prime}
		p_1' = x_2 - x'_1 \hspace{1cm} p_n' = x_1'-x_n 
\end{equation}
\ni whereas the region momenta of the resulting $(n+1)$-particle leading singularity satisfy
\begin{equation}
\label{unprime}
		p_1 = x_2 - x_1 \hspace{1cm} p_n = x_1-x_n 
\end{equation}
\ni with similar expressions for the fermionic components. Subtracting equations~\eqref{prime} and ~\eqref{unprime} we find two expressions for the shifted region momentum

\begin{eqnarray}
\label{Dualdeform}
		{x_1'} ^{\al\dal} &=& x_1^{\al\dal} - \frac{\la n+1\,n\ra}{\la 1\,n \ra} \lb_1^{\al} \tlb_{n+1}^{\dal} \nn\\
		&=& x_{n+1}^{\al\dal} + \frac{\la n+1\,1\ra}{\la n\,1 \ra} \lb_n^{\al} \tlb_{n+1}^{\dal} .
\end{eqnarray}

\ni whose equality is guaranteed by momentum conservation and is found directly with an application of the Schouten identity. To express the inverse soft factor in correctly in region momenta we should rewrite $\tlb_{n+1}$ in terms of the primary variables, but we will now find more convenient variables for the inverse soft factor.

\subsection{Momentum Twistors}


We now explain how an inverse soft factor acts on leading singularities when expressed as functions of momentum twistors~\cite{Hodges:2009hk,Mason:2009qx} which are designed to manifest the dual superconformal symmetry of scattering amplitudes in $\cN=4$ SYM~\cite{Drummond:2008vq}. 

\begin{figure}[htp]
\centering
\includegraphics[height=6cm]{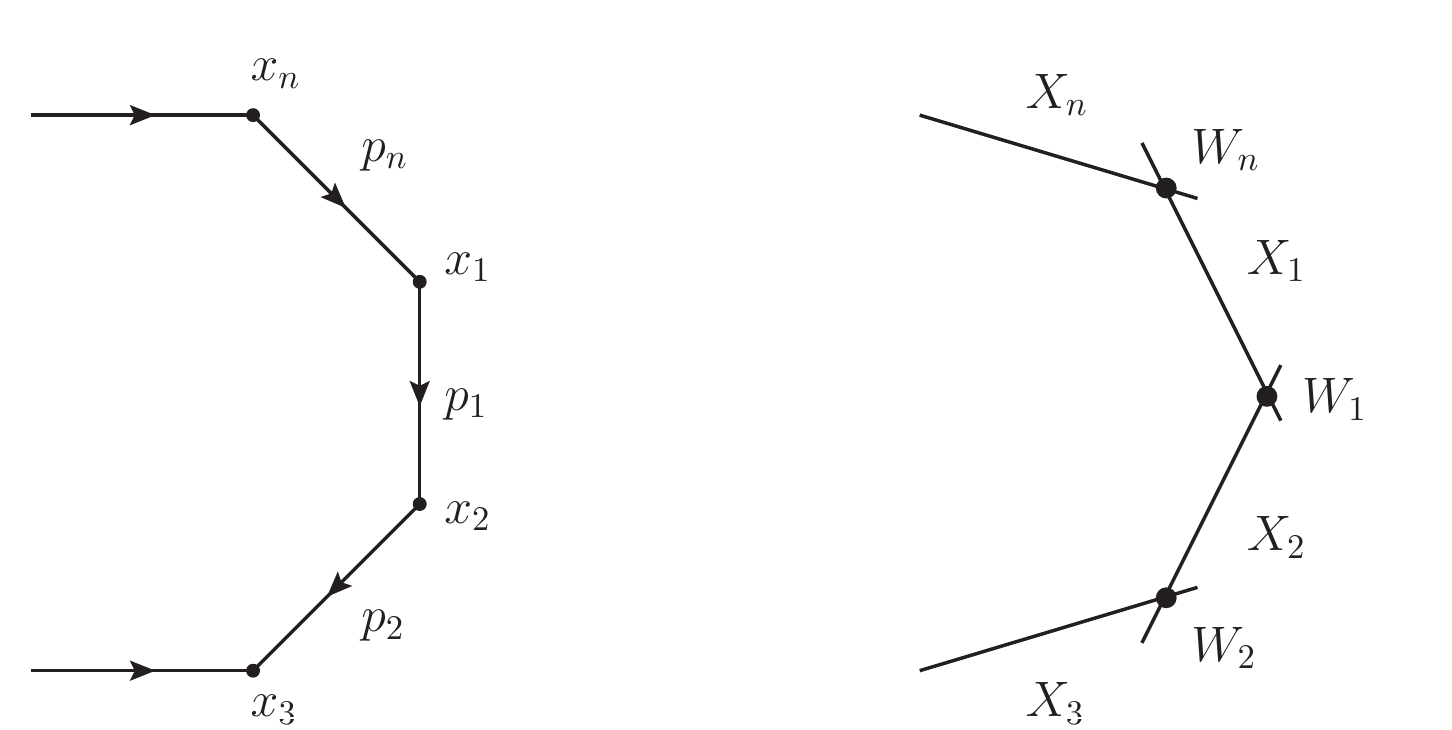}
\caption{\emph{The momentum twistor correspondence with dual momentum space.}}
\label{fig:Momtwistor}
\end{figure}

The cusps $x_i$ of the null polygon formed by the region momenta define lines $X_i$ in momentum twistor space through the standard twistor correspondence - see figure~\ref{fig:Momtwistor}. Since the cusps are null separated $
(x_i - x_{i-1})^2 = 0$, then adjacent lines $X_{i-1}$ and $X_i$ intersect and define momentum twistors $W_i$. The spinor components of the momentum twistor $W_i = (\lb_{i\al},\mu^{\dal}_i,\chi^A_i)$ are then defined by the incidence relations that simply state the the momentum twistor $W_i$ is incident on the point $x_i$ in region momentum space,
\begin{eqnarray}
\label{Components1}
		\mu^{\dal}_i = -ix^{\al\dal}_i\lb_{\al i} \qquad
		\chi^A_i = -i\theta^{A\al}_i\lb_{\al i}
\end{eqnarray}

\ni The components of the momentum twistors are also directly related to the original momentum superspace variables by the following 

\begin{eqnarray}
\label{Components2}
		i\tlb_i = \frac{ \la i-1,i \ra \mu_{i+1} + \la i+1,i-1 \ra \mu_{i} + \la i,i+1 \ra \mu_{i-1}}{\la i-1,i \ra\la i,i+1 \ra} \nn\\
		i\eta_i = \frac{ \la i-1,i \ra \chi_{i+1} + \la i+1,i-1 \ra \chi_{i} + \la i,i+1 \ra \chi_{i-1}}{\la i-1,i \ra\la i,i+1 \ra}.
\end{eqnarray}

Now consider adding particle $n+1$ with an inverse soft factor. Since only the dual coordinates $(x_1,\theta_1)$ are shifted then only the line $X_1$ is changed in momentum twistor space. However, from equation~\eqref{Dualdeform} we find that the momentum twistor $W_1$ is unchanged

\begin{eqnarray}
		{\mu_1'}^{\dal} &=& -i{x'_1}^{\al\dal}\lb_{1\al}  = -ix_1^ {\al\dal} \lb_{1\al} = \mu_1^{\dal}
\end{eqnarray}

\begin{figure}[htp]
\centering
\includegraphics[height=5.5cm]{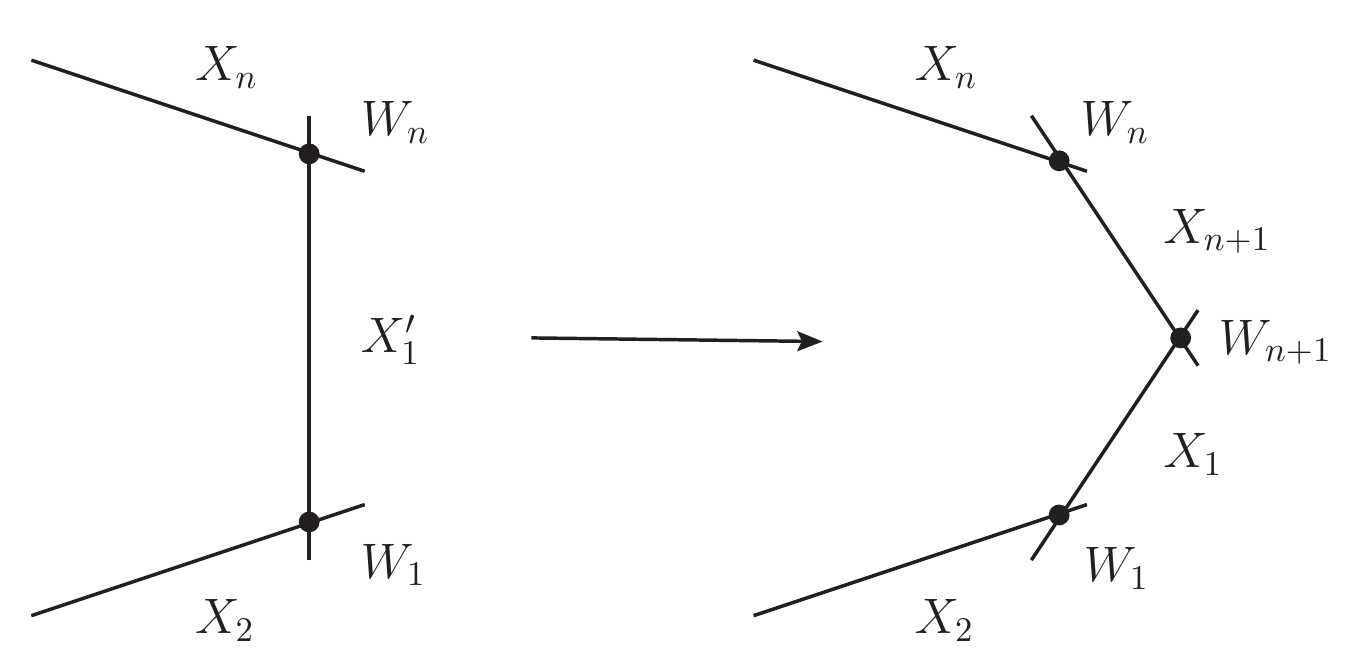}
\caption{\emph{Adding particle $\{n+1\}$ with an inverse soft factor in momentum twistor space.}}
\label{fig:ISLmomtwistor}
\end{figure}

\ni and so none of the momentum twistors are shifted in the inverse soft factor. The inverse soft factor simply shifts the line $X_1$ and adds an additional line $X_{n+1}$ together with the momentum twistor $W_{n+1}$ formed from their intersection - see figure~\ref{fig:ISLmomtwistor}.

Leading singularities are rational functions of the momentum twistors and once the MHV superamplitude has been removed they have weight zero in each momentum supertwistor. Consider then adding particle $(n+1)$ to a leading singularity $R_n(W_1,\dots,W_n)$ that depends on some subset of the momentum twistors $\{W_1,\ldots,W_n \}$. Then the resulting leading singularity $R_{n+1}(W_1,\ldots,W_n)$ is the same function of the original momentum twistors,

\begin{equation}
		R_{n+1}(W_1,\ldots,W_n) \equiv R_{n}(W_1,\ldots,W_n).
\end{equation}

\ni The new leading singularity does not depend on the new momentum twistor $W_{n+1}$. However, the relationship between the momentum twistors and the external momenta has been changed non-trivially.  In the opposite direction, soft limits are equally as simple in momentum twistor space. A leading singularity has non-zero soft limit $\lb_i\rightarrow 0$ when it does not depend on the momentum twistor $W_i$. For example, when particle $n$ becomes soft we have

\begin{equation}
\cR(W_1,\ldots,W_{n-1}) \longrightarrow \cR(W_1,\ldots,W_{n-1}) \hspace{0,5cm} \mathrm{as} \hspace{0.5cm} \lb_n\longrightarrow 0\, ,
\end{equation}

\ni so that the result of the soft limit is the same function of original momentum twistors.


Dual superconformal transformations act linearly on superamplitudes written in terms of momentum twistors. The standard superconformal generators are then the level-one generators of the Yangian algebra. Once the tree-level MHV superamplitude has been removed, the Yangian generators written in terms of momentum twistors are~\cite{Drummond:2010qh}

\begin{equation}
{j_{\cA}}^{\cB} = \sum\limits_{i=1}^n W_{i\cA} \frac{\partial}{\partial W_{i\cB}}  
\end{equation}
\begin{equation}
{{j^{(1)}}_{\cA}}^{\cB} = \sum\limits_{i< j}(-1)^{\cC}\left[ W_{i\cA} \frac{\partial}{\partial W_{i\cC}}  W_{j\cC} \frac{\partial}{\partial W_{j\cB}}	- (i\leftrightarrow j)\right]
\end{equation}

\ni It has been shown that the grassmannian formula written in momentum twistor variables transforms as a total derivative under the generators, and therefore all residues of the grassmannian formula are Yangian invariant~\cite{Drummond:2010qh}. Since it is conjectured that all leading singularities are residues of the grassmannian formula then we would expect the inverse soft factor to be a Yangian invariant operation. It is straightforward to check by acting with the level-one generators directly that this is indeed the case.



\section{Leading Singularities}
\label{LeadingSingularities}

A leading singularity of an $l$-loop amplitude is specified by $4l$ propagators going on-shell and a solution to the resulting cut conditions. Leading singularities are then associated with channel diagrams, having $4l$ lines denoting cut propagators and $(3l+1)$ vertices indicating tree amplitudes in the resulting factorisation~\cite{Cachazo:2008dx,Cachazo:2008vp,Cachazo:2008hp}. Primitive leading singularities are those containing only MHV and $\MHVbar_3$ vertices and correspond to individual grassmannian residues~\cite{Bullimore:2009cb}. Any leading singularity may then be expressed as a sum of primitive leading singularities by repeated BCFW expansion of the vertices. Therefore in the following we focus exclusively on primitive channel diagrams.


\subsection{Generalised Unitarity in Twistor Space}

\begin{figure}[h]
\centering
\includegraphics[height=25mm]{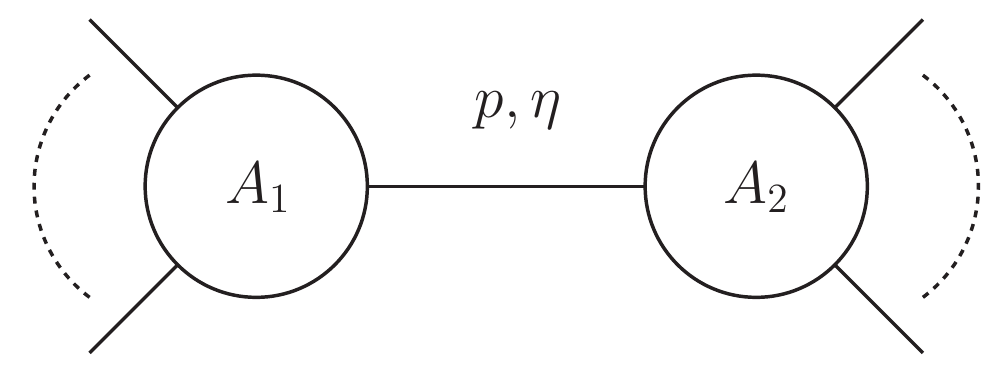}
\caption{\emph{The resulting factorisation when an internal propagator goes on-shell.}}
\label{Fig:InnerProduct}
\end{figure}

Consider computing the residue of a loop amplitude when one of its internal propagators goes on-shell. Only Feynman diagrams containing this propagator contribute to the residue and standard LSZ arguments ensure that the residue is the product of two subamplitudes on either side of the cut, summed over all possible internal states - see figure~\ref{Fig:InnerProduct}. This calculation is building block for all generalised unitarity calculations.

In $\cN=4$ SYM the sum over particles and helicities may be replaced by a grassmann integral and the tree amplitudes by superamplitudes with external states labelled in on-shell superspace~\cite{Drummond:2008bq,ArkaniHamed:2008gz}. The unitarity cut then becomes (see figure~\ref{Fig:InnerProduct})

\begin{equation}
\label{Lightconeintegral}
\oint \frac{\rd^4p}{p^2}\, \rd^4\eta \ A_1(\ldots,\{p,\eta\})\, A_2(\{-p,\eta\},\ldots)\ ,
\end{equation}

\ni where the contour $|p^2|=\varepsilon$ constrains the integral to the null cone. In split signature, this may be transformed directly into twistor space (not to be confused with momentum twistor space) via a half Fourier Transform~\cite{Witten:2003nn} leading to the standard twistor inner product~\cite{Bullimore:2009cb}
\begin{equation}
\label{Twistorinnerproduct}	
		 \int \rD^{3|4}W\, A_1(\ldots, W)\, A_2(W,\ldots)\ .
\end{equation}

\ni This simple form of the unitarity cut in twistor space allows an imediate translation between the channel diagram and the twistor space support of leading singularities~\cite{Bullimore:2009cb,Kaplan:2009mh}. This dictionary is summarised below in figure~\ref{fig:Twistorsupportrules}. 

\begin{figure}[htp]
\centering
\includegraphics[height=11cm]{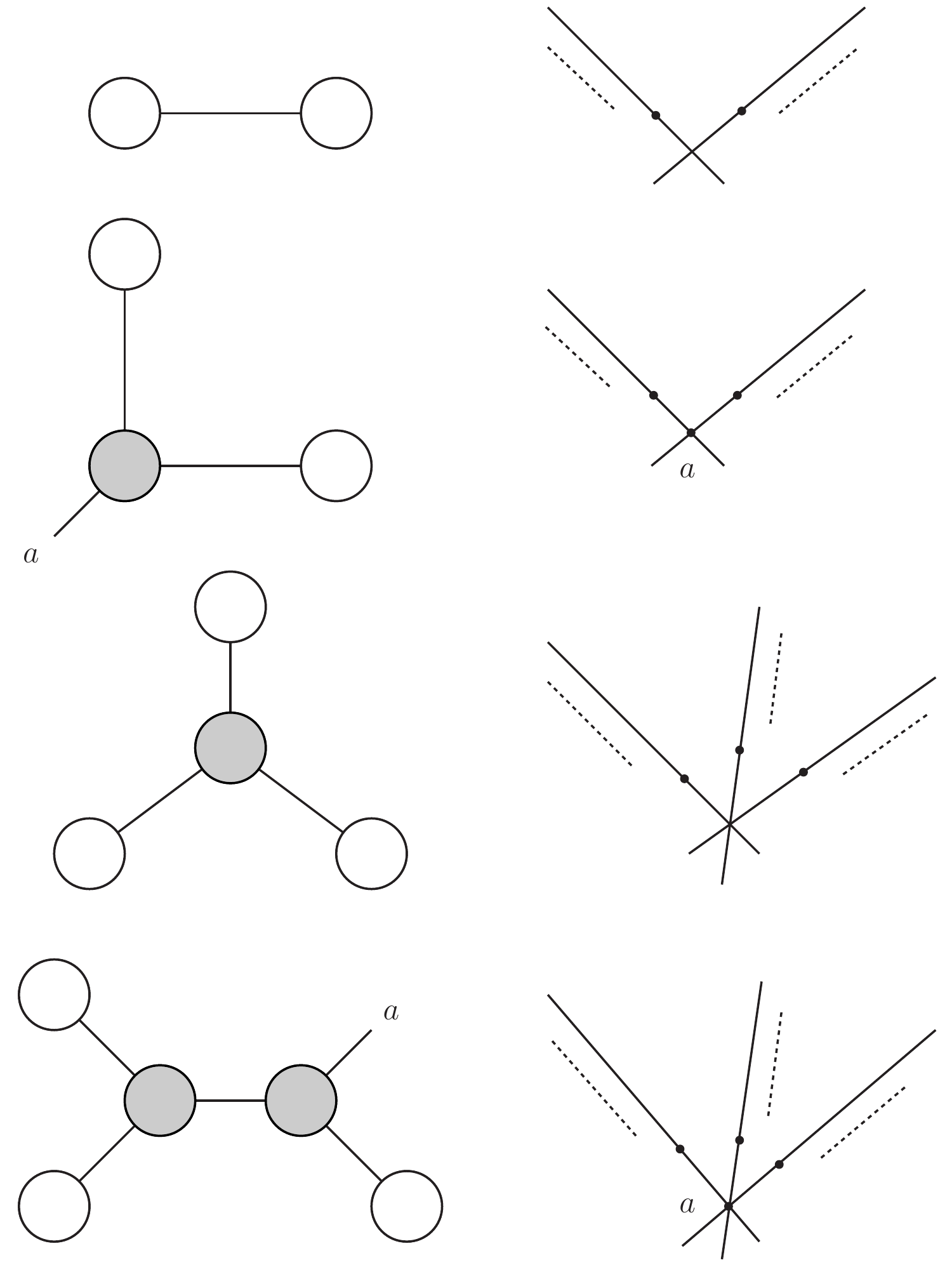}
\caption{\emph{The dictionary between the twistor space support of primitive leading singularities and their channel diagrams.}}
\label{fig:Twistorsupportrules}
\end{figure}


\subsection{Inverse Soft Factors and Channel Diagrams}

\begin{figure}[h]
\centering
\includegraphics[height=5.5cm]{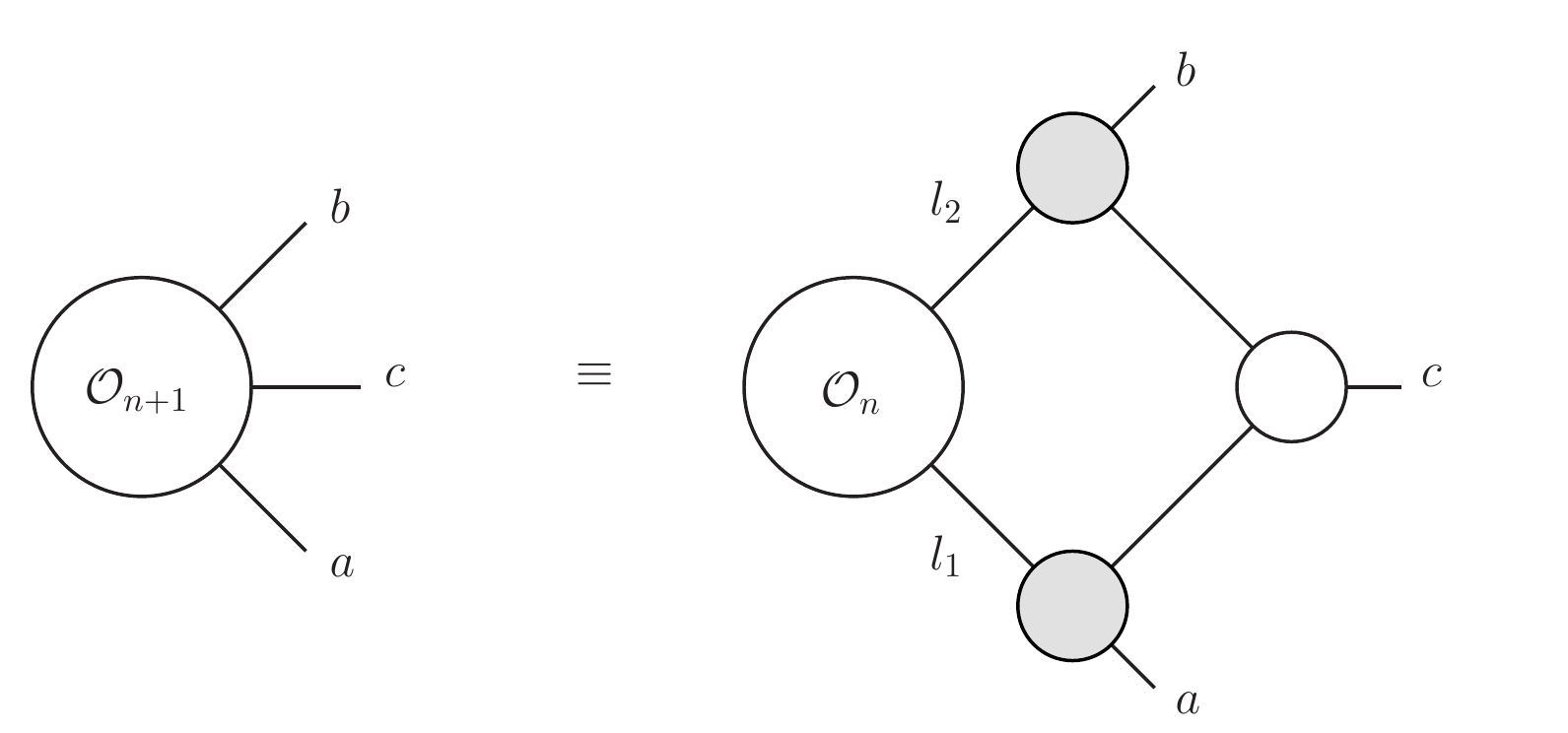}
\caption{\emph{The action of an inverse soft factor on the channel diagrams of a leading singularity}}
\label{fig:ISLchannel}
\end{figure}

The inverse soft factor may be understood as a simple application of generalised unitarity. Adding the particle $c$ to the leading singularity $\cO_n(a,b,\dots)$ then the result $\cO_{n+1}(a,c,b,\ldots)$ corresponds simply to the channel diagram shown in figure~\ref{fig:ISLchannel}. 

This result may be demonstrated with generalised unitarity in on-shell superspace~\cite{Drummond:2008bq}. The solution of the cut conditions for the on-shell loop momenta in $\cO_n(l_1,l_2,\ldots)$ have been written down explicitly when at least one corner is massless in~\cite{Berger:2008sj}.  In the spinor notation the relevant solution becomes (see notation in figure~\ref{fig:ISLchannel})
\begin{eqnarray}
\label{loopmomenta}
l_1 &=& \lb_a\left( \tlb_a + \frac{\la bc\ra}{\la ba \ra}\tlb_c \right) \nn\\
l_2 &=& \lb_b\left( \tlb_b + \frac{\la ac\ra}{\la ab \ra}\tlb_c \right) \,.
\end{eqnarray}

\ni Performing the grassmann integrations and evaluating on the above solution to the cut conditions, the corresponding grassmann parameters $\eta_{l_a}$ and $\eta_{l_a}$ are determined to be
\begin{eqnarray}
\label{loopspinors}
\eta_{l_1} &=&  \eta_a + \frac{\la bc\ra}{\la ba \ra}\eta_c  \nn\\
\eta_{l_2 }&=&  \eta_b + \frac{\la ac\ra}{\la ab \ra}\eta_c  \,.
\end{eqnarray}

\ni Finally the three-particle superamplitudes in figure~\ref{fig:ISLchannel} turn into the correct MHV factor in equation~\eqref{ISLdef} so that we recover the required result

\begin{eqnarray}
\frac{\la ab \ra}{\la ac \ra\la cb \ra} \cO(\l_1,\eta_{l_1};l_2,\eta_{l_2})
\end{eqnarray}

\ni where the loop momenta and grassmann parameters are given in equations~\eqref{loopmomenta} and~\eqref{loopspinors}. We will use this interpretation of the inverse soft limit extensively in the following to identify grassmannian residues with leading singularities and find expressions for them in terms of momentum twistors.


\begin{figure}[h]
\centering
\includegraphics[height=4cm]{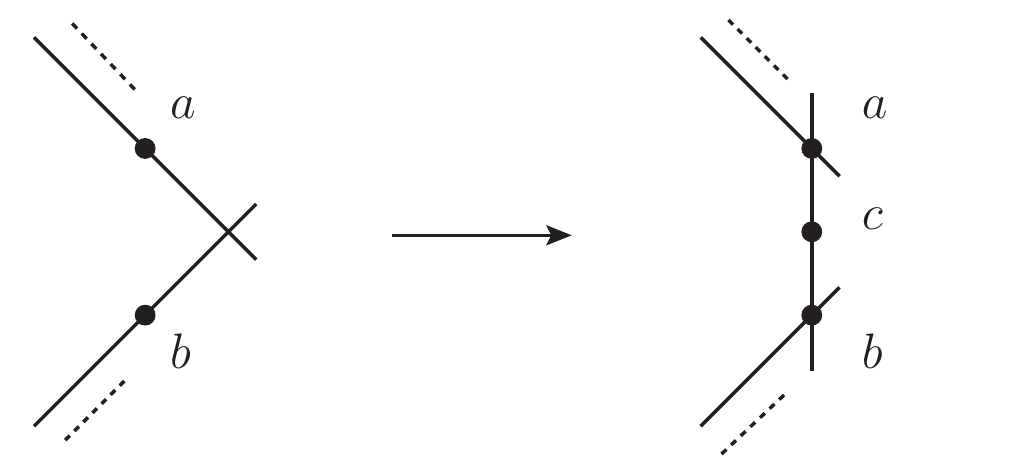}
\caption{\emph{The action of an inverse soft factor on the twistor space support.}}
\label{fig:ISLtwistor}
\end{figure}

Translating the statements about channel diagrams to those about twistor support, we find that the result of an inverse soft factor is supported where the additional twistor $c$ is collinear with $a$ and $b$. Figure~\ref{fig:ISLtwistor} indicates the twistor support when particle c is added between particles $a$ and $b$ on adjacent MHV vertices. In general, whenever three points $\{a-1,a,a+1\}$ are collinear in twistor space, then the leading singularity is independent of the momentum twistor $W_a$ in momentum space in agreement with the action of an inverse soft limit.


\begin{figure}[h]
\centering
\includegraphics[height=9.0cm]{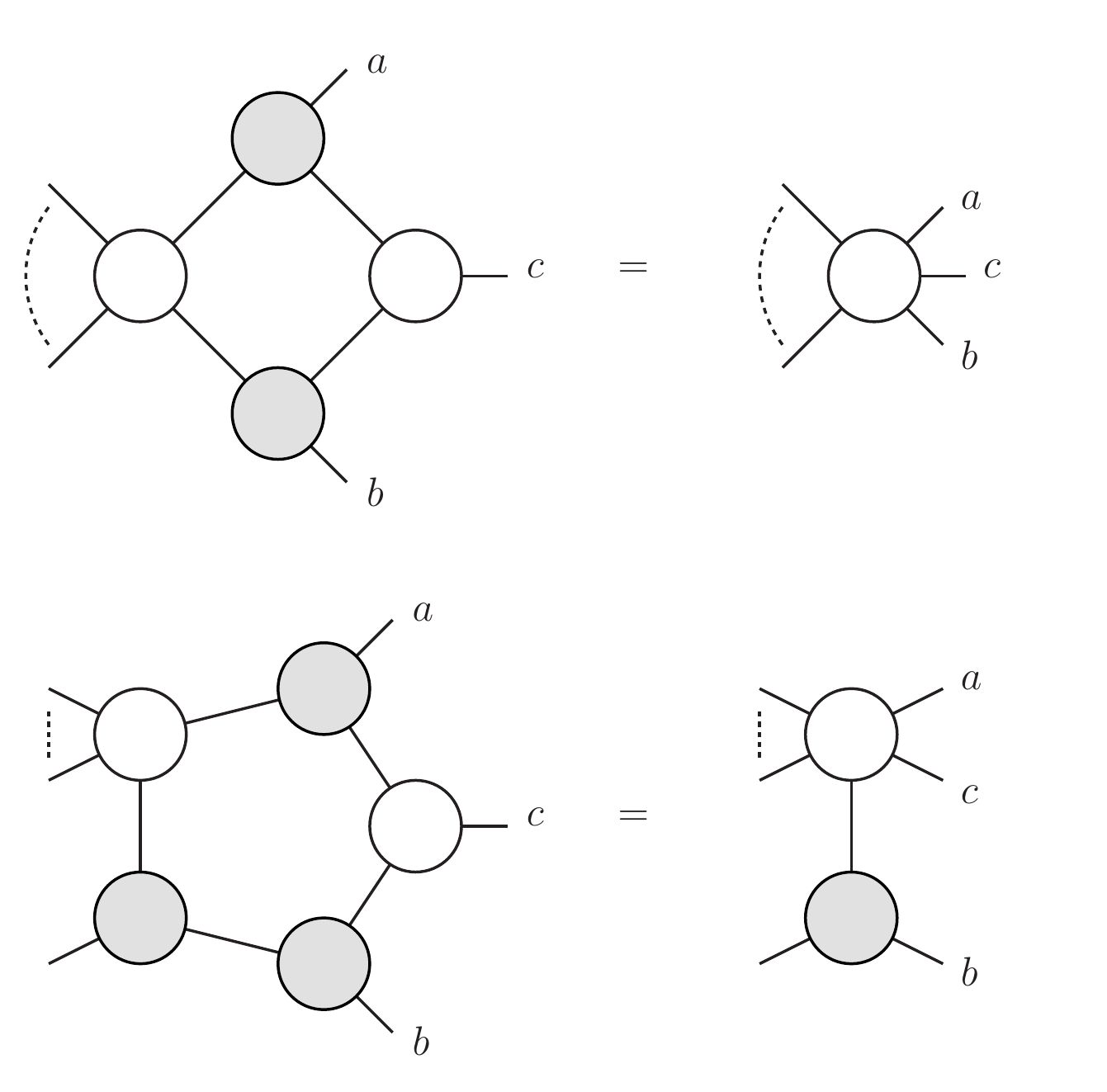}
\caption{\emph{The channel diagrams corresponding to inverse soft factors that simply add more particles to an existing MHV vertex.}}
\label{fig:ISLMHV}
\end{figure}

In many cases an inverse soft factor simply adds another particle to an existing MHV vertex without changing the sturcture of the channel diagram. We have already seen that adding particle $c$ between two legs $a$ and $b$ of the same MHV vertex adds an additional particle to that vertex. Similarly, adding $c$ between particle $a$ on an MHV vertex and particle $b$ on an adjacent $\MHVbar_3$ vertex again adds an additional particle to the MHV vertex - see figure~\ref{fig:ISLMHV}.



\section{Grassmannian Residues}

Having discussed the action inverse soft limits on leading singularities, we now turn to the application of identifying primitive leading singularities with residues of the grassmannian integral~\cite{ArkaniHamed:2009dn}. For N$^{k-2
}$MHV superamplitudes with $n$ particles, the formula is a contour integral in the grassmannian $G(k,n)$. When external states are transformed to dual twistor space (where MHV amplitudes are supported on a line in our conventions) we have

\begin{equation}
\label{GrassmannianW}
A(W_1,\ldots, W_n) = \frac{1}{\mathrm{vol}\,GL(2)} \int \frac{\rd^{k\times n}C}{(1)\ldots(n)} \prod_{r=1}^k d^{4|4}Y_r \prod_{i=1}^{n} \delta^{4|4}(W_i - C_{ri}Y_r)\, ,
\end{equation}
\vspace{0.5mm}

\ni where $(i)=(i,\ldots,i+k)$ are minors of the $(k\times n)$-matrix $C_{ri}$ of homogeneous coordinates on $G(k,n)$. Local residues are defined by $(k-2)(n-k-2)$ conditions on the minors and as we will discuss correspond to projective configurations of $n$ points in $\CP^{k-1}$ with certain localisation properties.


\subsection{Projective Geometry in the Grassmannian}

Projective geometry in the grassmannian plays an important role in understanding the individual residues of the grassmannian integral~\cite{ArkaniHamed:2009dg}. Here we would like to understand how the geometry of projective configurations arises.  

Since superamplitudes in $\cN=4$ are invariant under little group transformations which rescale the twistors~\cite{ArkaniHamed:2008gz} then the grassmannian formula~\eqref{GrassmannianW} is invariant provided the homogeneous coordinates transform in addition
\begin{equation}
W_i \longrightarrow t_iW_i \hspace{1cm} C_{r i} \longrightarrow t_i\, C_{r i}.
\end{equation}

\ni Therefore the little group transformations form a subgroup of the global $GL(n)$ symmetry of the grassmannian formula
\begin{equation}
 W_i \longrightarrow {L_i}^jW_j  \hspace{1cm}  C_{ri} \longrightarrow {L_i}^j C_{rj} \, .
\end{equation}

\ni Since multiples of the identity act trivially, there is a free action of $H=(\C^*)^n / \C^*$ and we may pass to the quotient $G(k,n)\, /\, H$. On the other hand, since columns can be rescaled separately, the homogeneous coordinates  $C_{r i}$ define a configuration of $n$ points in the projective space $\CP^{k-1}$. However, there is in addition a local $GL(k)$ symmetry
\begin{equation}
C_{r i} \longrightarrow {\Lambda_r}^s\, C_{s i}
\end{equation}

\ni which acts projectively on the configurations of points. Therefore the equivalence classes of projective configurations under $GL(k)$ match up with the grassmannian $G(k,n)$ modulo little group transformations. This is called the Gelfand-MacPherson correspondence~\cite{Gelfand82},
\begin{equation}
		G(k,n)\, /\, H = (\bP^{k-1})^n\, /\, GL(k).
\end{equation}

The characteristic properties of grassmannian residues are therefore properties of projective configurations of $n$-points in $\CP^{k-1}$ that are invariant under local $GL(k)$ transformations.  Such properties are exactly localisation properties. 


\subsection{Grassmannian Localisation and Twistor Support}

The grassmannian localisation on curves of degree $(k-1)$ in $\CP^{k-1}$ has been important in understanding tree amplitudes and the connection to twistor string theory~\cite{ArkaniHamed:2009dg,Nandan:2009cc,Bourjaily:2010kw}. However, here we focus on individual residues and localisation on planes of various codimension.

The grassmannian localisation of $n$ points in projective space $\CP^{k-1}$ is related to support in twistor space $\CP^3$ through the delta-functions in the grassmannian integral~\cite{Bullimore:2009cb,ArkaniHamed:2009sx}

\begin{equation}
	\int \prod\limits_{r=1}^{k} \rd^{4|4}Y_r \prod\limits_{i=1}^n \delta^{4|4}(W_i - C_{r i}Y_r).
\end{equation}

\ni Immediately there are constraints on the twistor space support arising purely from the dimension of the space $\CP^{k-1}$. In particular MHV amplitudes are supported on a line and NMHV amplitudes on a plane in twistor space.

Further grassmannian localisation in $\CP^{k-1}$ is tested by the vanishing of the minors of $G(k,n)$; when the minor $(i_1,\ldots,i_k)$ vanishes then the points $\{i_1,\ldots,i_k\}$ lie in a codimension one subspace in $\CP^{k-1}$. For NMHV and N$^2$MHV amplitudes this translates directly into statements about the twistor space support:
\pagebreak
\begin{itemize}
\item NMHV: $(i_1\,i_2,\i_3)=0$ implies that the points $\{i_1,i_2,i_3\}$ are collinear in twistor space.
\item N$^2$MHV: $(i_1\,i_2\,i_3\,i_4)=0$ implies that the points $\{i_1,i_2,i_3,i_4\}$ are coplanar in twistor space.
\end{itemize}

\ni However, for $\mathrm{N}^3\mathrm{MHV}$ amplitudes and higher the relation between grassmannian localisation and twistor support is less immediate and here we almost exclusively on NMHV and $\Nsq$ amplitudes.


\subsection{The Momentum Twistor Grassmannian}

In~\cite{Mason:2009qx} an equivalent grassmannian integral in $G(k-2,n)$ has been found where the natural variables are momentum twistors. Once the overall MHV superamplitude has been cleared the integral becomes (denoting $p=k-2$)
 
\begin{equation}
\label{Momgrassmannian}
		R(W_1,\ldots, W_n) = \frac{1}{\mathrm{vol}\,GL(p)} \int \frac{\rd^{n\times p}D}{(D_1\ldots D_p)\ldots(D_n\ldots D_{p-1})} \prod_{r=1}^{p} \delta^{4|4}(D_{ri}W_i) \, .
\end{equation}
\vspace{0.5mm}

\ni where $(D_i \dots D_{i+p})$ are minors of the matrix $D_{ri}$ of homogeneous coordinates on $G(k-2,n)$. Individual residues are again defined by $(k-2)(n-k-2)$ conditions on the minors. The minors $(D_i \ldots D_{i+p})$ are directly proportional to the minors $(i)$ of the original $G(k,n)$ grassmannian upt to kinematic factors~\cite{ArkaniHamed:2009vw} so there is a clear correspondence between the residues of the two grassmannian integrals. 

The original $G(k,n)$ grassmannian integral is naturally written with external states transformed to twistor space and therefore manifests the standard superconformal symmetry. It is also more suited for the geometric interpretation and identification of residues. On the other hand, the $G(k-2,n)$ grassmannian is written in terms of momentum twistors and manifests dual superconformal invariance. The smaller minors means that it is also simpler to calculated individual residues. In fact individuals residues are invariant under the Yangian of the superconformal (or dual superconformal) algebra~\cite{Drummond:2010qh,Drummond:2009fd} and the Yangian symmetry determines uniquely the correct integrand of the grassmannian integral~\cite{Drummond:2010uq}.



\section{NMHV Amplitudes}

For NMHV amplitudes the momentum twistor grassmannian is $G(1,n) = \CP^n$ and the integral formula~\eqref{Momgrassmannian} becomes

\begin{equation}
\label{NMHVmomgrass}
		R(W_1,\ldots,W_n)= \frac{1}{\mathrm{vol}(\C^*)} \int \frac{\rd^n D}{D_1\ldots D_n}\, \delta^{4|4}(D_i W_i)\, .
\end{equation}
\vspace{0.1mm}

\ni The homogeneous coordinates $D_i$ on projective space $\CP^n$ correspond directly to the minors $(i)$ of the $G(3,n)$ grassmannian. Residues are then defined by the vanishing of $(n-5)$ coordinates $D_i$ and are labelled by the corresponding minors or more conveniently here by the five corresponding minors $\overline{(i_1)\ldots(i_5)}$ that do not vanish.


All residues are automatically supported on a plane in twistor space. When the $(i)=0$ there is grassmannian localisation with $\{i-1,i,i+1\}$ collinear in $\CP^2$. The residue then has support when twistors $\{i-1,i,i+1\}$ are collinear. From the delta functions in equation~\eqref{NMHVmomgrass} it is clear the residue is then independent of momentum twistor $W_i$.


\subsection{Inverse Soft Factors}

When particle $i$ is added in between $i-1$ and $i+1$ by an inverse soft factor then the resulting residue is supported where the twistors $\{i-1,i,i+1\}$ are collinear. Therefore the minor $(i)$ vanishes on this residue and it is independent of momentum twistor $W_i$ in agreement with the general arguments made in section~\ref{InverseSoftLimits}. The inverse soft factor then has a simple action on residues; adding the particle $j$ to the residue $(i_1)\ldots(i_{n-5})$ then we have
\begin{equation}
		(i_1)\ldots(i_{n-5}) \longrightarrow (i_1)\ldots(i_{n-5})(j)
\end{equation}

\ni or equivalently labelling residues by the minors that do not vanish then $\overline{(l_1)\ldots(l_5)}  \longrightarrow \overline{(l_1)\ldots(l_5)}$ where the set $\{l_1,\dots,l_5\}$ is the complement of $\{i_1,\ldots,i_{n-5}\}$ in $\{1,\ldots,n\}$.


\subsection{Residues}

The individual residues of the NMHV grassmannian integral may all be evaluated from the momentum twistor integral with the result~\cite{Mason:2009qx}
\begin{equation}
		\overline{(i)(j)(k)(l)(m)} = R(i,j,k,l,m)\, .
\end{equation}

\ni The quantity $R(i,j,k,l,m)$ is the basic dual superconformal invariant which is a homogeneous function of five momentum twistors~\cite{Mason:2009qx}. It is antisymmetric under interchange of any two of its arguments and being a residue of the grassmannian formula is Yangian invariant. This Yangian invariant forms the basic building block for superamplitudes of any degree in $\cN=4$ SYM.

\begin{figure}[htp]
\centering
\includegraphics[height=4.5cm]{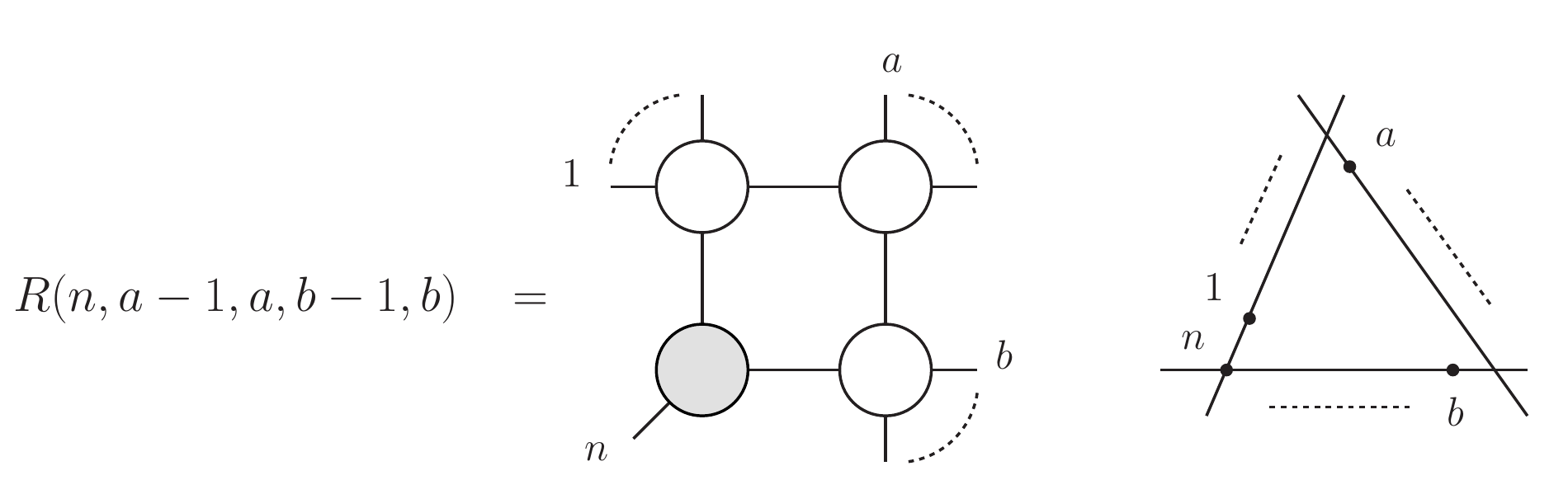}
\label{fig:NMHVbasic}
\caption{\emph{The channel diagram and twistor support of the basic R-invariant $R(n,a-1,a,b-1,b)$.}}
\end{figure}

The simplest case is when two pairs of minors are adjacent. In this case the grassmannian residue becomes the standard dual superconformal invariant $R(n,i-1,i,j-1,j) \equiv R_{n;ij}$ appearing in BCFW expansions of the NMHV tree superamplitude~\cite{Drummond:2008cr,Britto:2005fq}, for example,

\begin{equation}
\label{NMHVtreeamplitude}
		A^{\NMHV}(1,\ldots,n) = A^{\MHV}_n \times \sum\limits_{1 < i,j < n} R_{n;ij} \, .
\end{equation}

\ni where in such sums $i$ and $j$ must be separated by at least two. Individually, the residues are one-loop leading singularities (see figure~\ref{fig:NMHVbasic}) reflecting the original derivation of the BCFW recursion relations from the IR consistency of one-loop amplitudes~\cite{Britto:2004ap}.

\begin{figure}[htp]
\centering
\includegraphics[height=5cm]{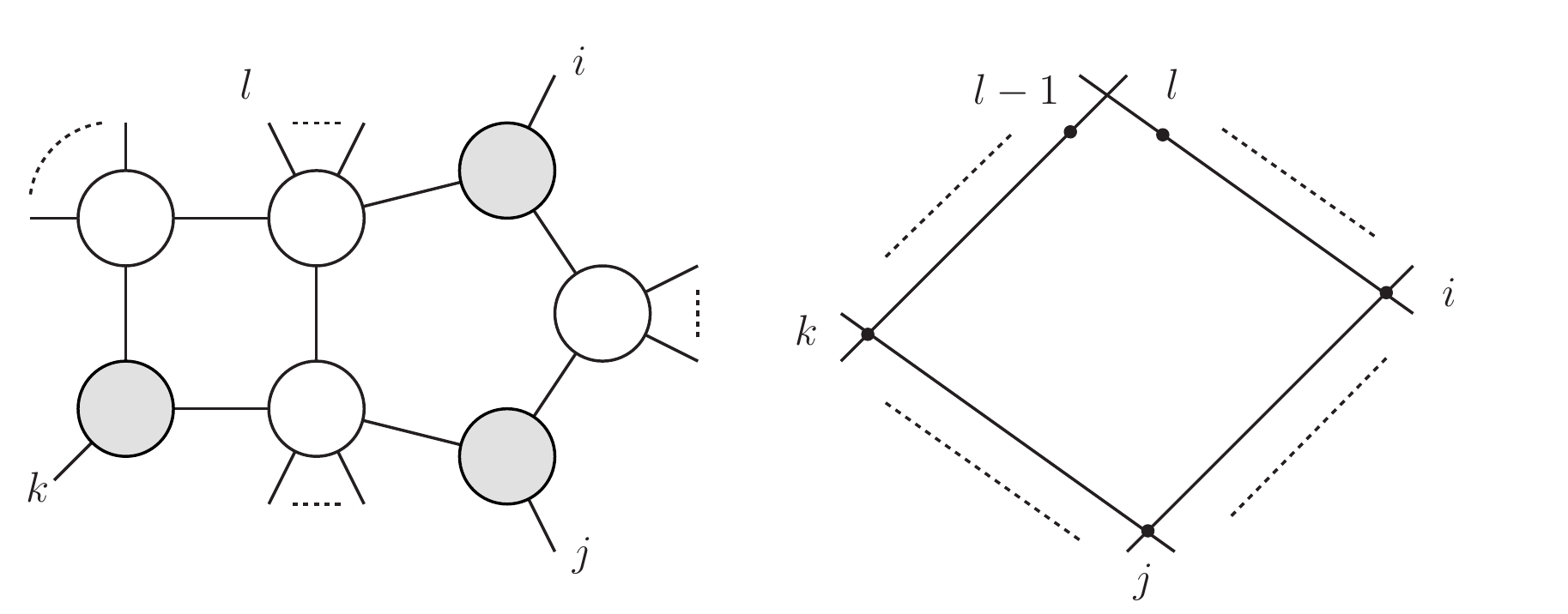}
\caption{\emph{A channel diagram corresponding to the Yangian invariant $R(i,j,k,l-i,l)$.}}
\label{fig:NMHVpentabox}
\end{figure}
\begin{figure}[htp]
\centering
\includegraphics[height=6cm]{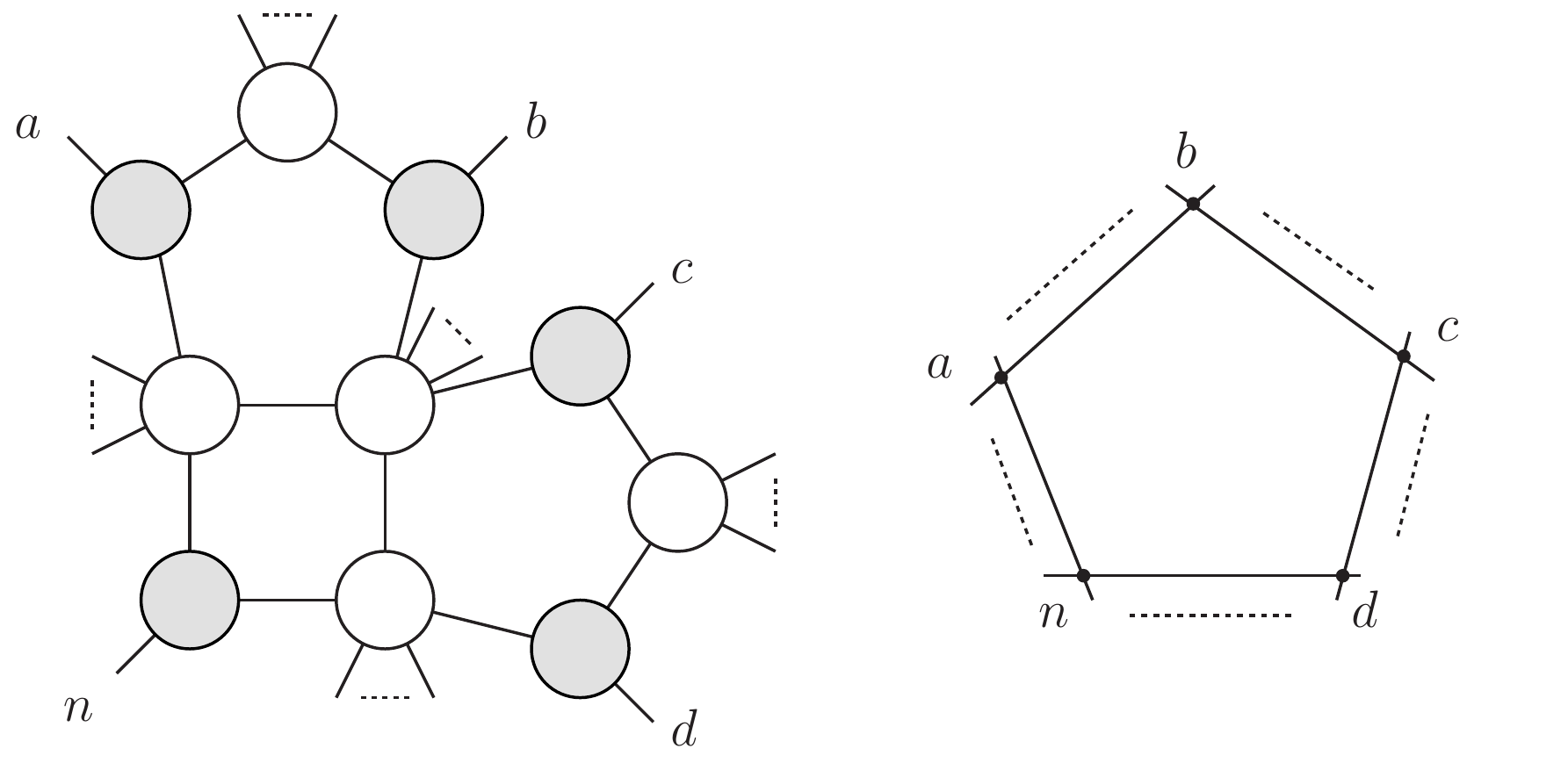}
\caption{\emph{A channel diagram corresponding to the Yangian invariant $R(i,j,k,l,m)$.}}
\label{fig:NMHVgeneric}
\end{figure}

In the more generic cases $R(i,j,k,l-1,l)$ and $R(i,j,k,l,m)$ the residues correspond to two-loop and three-loop leading singularities respectively~\cite{Bullimore:2009cb}. The channel diagrams are constructed from the twistor support which is turn immediate from the vanishing minors - see figures~\ref{fig:NMHVpentabox} and~\ref{fig:NMHVgeneric}. It is clear from the dependence on momentum twistors that the generic residues may all be constructed by inverse soft factors. The action on the channel diagrams then agrees with the general arguments in section~\ref{LeadingSingularities}.

\ni This exhausts all residues of the grassmannian integral and therefore we expect that there are no further Yangian invariants relevant for NMHV amplitudes. In general it is conjectured that there are no new Yangian invariants corresponding to primitive leading singularities past $3p$-loops for $\Np$ amplitudes~\cite{Bullimore:2009cb}.



\section{N$^2$MHV Amplitudes}

For $\Nsq$ amplitudes the momentum twistor grassmannian is $G(2,n)$ and after the MHV superamplitude has been cleared the grassmannian integral formula becomes

\begin{equation}
		R(1,\ldots,n)= \frac{1}{\mathrm{vol}(GL(2))} \int \frac{d^{2n} D}{(D_1D_2)\ldots (D_nD_1)}\, \prod\limits_{r=1}^2 \delta^{4|4}(D_{ri} W_i)\, .
\end{equation}
\vspace{0.1mm}

\ni The $2\times2$ minors $(D_i D_{i+1})$ are proportional to the minors $(i)=(i-1,i,i+1,i+2)$ of the $G(4,n)$ grassmannian which are labelled by their second column. Individual residues may now be standard and composite~\cite{ArkaniHamed:2009dn,ArkaniHamed:2009dg}. Standard residues are defined by $(2n-12)$ minors vanishing to first order, while composite residues by fewer than $(2n-12)$ minors with some vanishing to second order. 

Individual residues are determined by their grassmannian localisation in $\CP^3$ which translates directly into statements about twistor space support of the corresponding leading singularity. The residues are specified by the vanishing minors $(a_1)\cdots(a_m)$ with $m\leq(2n-12)$ and by their collinear localisation~\cite{ArkaniHamed:2009dg}. In addition it is useful to label the coplanar localisation so that following~\cite{ArkaniHamed:2009dg} we have the subscript and superscript notation:

\begin{itemize}
\item $(\cdots)_{m} \Rightarrow  \{ m-1, m,m+1\}$ are collinear in $\CP^3$
\item $(\cdots)^{m} \Rightarrow $ the complement of particles $\{ m-1,m,m+1\}$ are coplanar in $\CP^3$.
\end{itemize}


A useful tool to understand the structure of vanishing minors is the factorisation of minors~\cite{ArkaniHamed:2009dg}. In order to see this consider the two adjacent minors $(2)\equiv(1234)$ and $(3)\equiv(2345)$ whose simultaneous vanishing require the coplanarity of the two sets points $\{1234\}$ and $\{2345\}$. This may happen in two ways; firstly the points $\{12345\}$ are all coplanar, or secondly the points $\{234\}$ are collinear. For eight particles, this factorisation is denoted by

\begin{equation}
\label{factorisation1}
		(2)(3) \Rightarrow \left\{
		\begin{array}{c}
		(2)(3)^{7}\\
		(2)(3)_{3}\\
		\end{array} \right.	
\end{equation}
\vspace{0.1mm}

\ni although for more particles, we cannot include the coplanarity label $(\ldots)^7$ in the case that the points $\{12345\}$ are collinear.

The generic $\Nsq$ residue is highly composite with many minors vanishing to second order. For example, we will examine the eight-particle residues of the form  $(1)(2)^2(3)_{2}^{7}$ where the notation $(2)^2$ means that this minor vanishes to second order. However, for large numbers of particles this notation becomes cumbersome and the information may be recovered from the collinear labels. Therefore we will omit such labels, denoting the above simply by $(1)(2)(3)_{2}^{7}$.


\subsection{Inverse Soft Limits}

Consider adding the particle $i$ is added in between $i-1$ and $i+1$ with an inverse soft factor. Then we have shown that the result has support where the points $\{i-1,i,i+1\}$ are collinear in twistor space. For $\Nsq$ amplitudes, this requires one further condition on each of the minors $(i-1)$ and $(i)$ in order to produce the required factorisation $(i-1)(i)_{i}$ in which $\{i-1,i,i+1\}$ are collinear. Therefore adding particle $j$ to a generic residue $(i_1)\ldots(i_k)_{j_1,\ldots ,j_m}^{l_1,\ldots,l_n}$ with an inverse soft factor produces the following residue

\begin{equation}
		(i_1)\ldots(i_k)_{j_1,\ldots, j_m}^{l_1,\ldots,l_n} \longrightarrow (i_1)\ldots(i_k)(j-1)(j)_{j_1,\ldots j_mj}^{l_i,\ldots,l_n}
\end{equation}

\ni Following the earlier convention, if the minor $(j-1)$ already vanishes to first order before the inverse soft factor then the label will not be repeated in the resulting residue.


\subsection{Residues}

We will not attempt to identify all $\Nsq$ residues with leading singularities, but instead examine classes of residues built from inverse soft limits,which illustrate important features and patterns. Hopefully this will then allow the reader to construct many further examples. However, we will identify the leading singularities appearing in BCFW expansions of the tree superamplitude, for example~\cite{Drummond:2008cr}

\begin{equation}
	A_n^{\Nsq} = A_n^{\MHV} \sum_{1<i<j<n}\!\!\! R_{n;ij}
	\left[\,\sum_{i<k<l\leq j}\!\!\! R^{ij}_{n;ji;kl} 
	+ \sum_{j\leq k<l<n}\!\!\!R^{ij}_{n;kl}\right]\, 
\label{NNMHVtree}
\end{equation} 

\ni in which each term is a two-loop primitive leading singularity~\cite{Bullimore:2009cb}. The notation used in equation~\eqref{NNMHVtree} is explained in~\cite{Drummond:2008cr} and also further in the following text. Note that in the boundary cases $l=j$ and $k=j$ the superscripts in equation~\eqref{NNMHVtree} denote how the dual superconformal invariants are modified in these cases. We will see that when written in momentum twistors the required modification for the boundary terms becomes much more transparent.


\subsubsection{The Standard BCFW Terms}
\label{BCFWstandard}


First we consider generic residues appearing in solutions of the BCFW expansion of the $\Nsq$ tree superamplitude written down in~\cite{Drummond:2008cr}. From their twistor space support~\cite{Korchemsky:2009jv} it is straightforward to show that these terms correspond to pentabox channel diagrams with generic numbers of legs~\cite{Bullimore:2009cb}. 

\begin{figure}[htp]
\centering
\includegraphics[height=10cm]{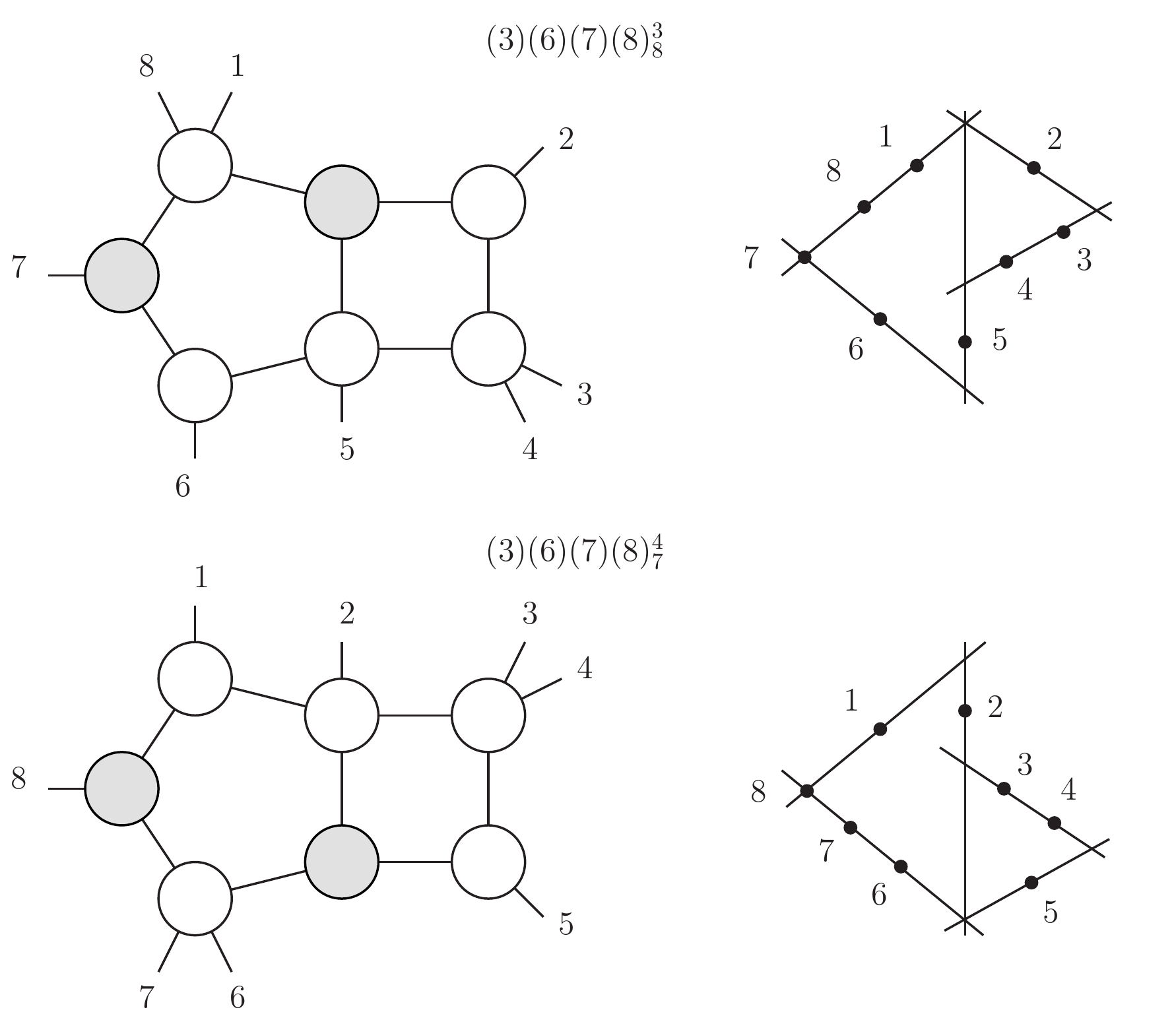}
\caption{\emph{The twistor support and channel diagrams of two standard BCFW terms.}}
\label{fig:8standard3678}
\end{figure}

Consider eight-particle residues where the three adjacent minors $(6)$,$(7)$ and $(8)$ vanish and the two pairs of adjacent minors factorise:

\[
		(6)(7) \Rightarrow \left\{
		\begin{array}{c}
		(6)(7)^{3}\\
		(6)(7)_{7}\\
		\end{array} \right.		
\]
\[
		(7)(8) \Rightarrow \left\{
		\begin{array}{c}
		(7)(8)^{4}\\
		(7)(8)_{8}\\
		\end{array} \right.	
\]\vspace{0.5mm}

\ni One possible residue is $(6)(7)(8)_{78}^{34}$ where the four points $\{6781\}$ are collinear and the minor $(7)$ vanishes to second order, however we will study this residue in the following section. We can define standard residues of the form $(6)(7)(8)_{7}^{4}$ and $(6)(7)(8)_{8}^{3}$ together with one more minor. In these cases the coplanar label follows automatically with the vanishing of the third minor. Here we consider the two residues $(3)(6)(7)(8)_{8}^{3}$ and $(3)(6)(7)(8)_{7}^{4}$. The twistor support is determined from the localisation and translated into channel diagrams - see figure~\ref{fig:8standard3678}. Expressions in terms of momentum twistors may be found by computing the residues explicitly~\cite{Mason:2009qx} or more easily by generalised unitarity with the result:
\begin{eqnarray}
\label{standard3678}
		(3)(6)(7)(8)_{8}^{3} &=& R(5,6,7,1,2)\,R(V,2,3,4,5) \hspace{1cm} V = \la 5,6,7,[1\ra 2] \nn\\
		(3)(6)(7)(8)_{7}^{4} &=& R(8,1,2,5,6)\,R(U,2,3,4,5) \hspace{1cm} U = \la 8,1,2,[5\ra 6]\, .
\end{eqnarray}
\begin{figure}[htp]
\centering
\includegraphics[height=5.5cm]{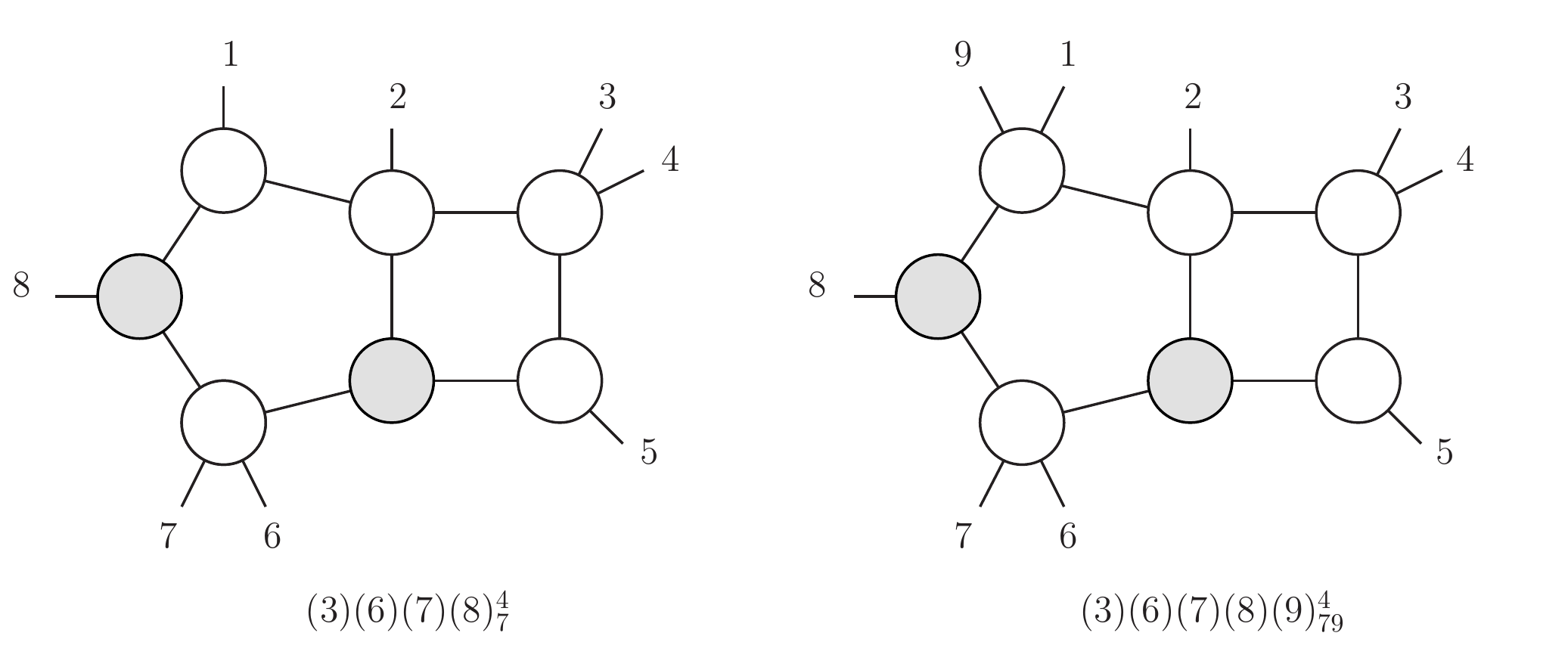}
\caption{\emph{BCFW residues with eight and nine particles related by an inverse soft factor.}}
\label{fig:ISLcomposite678}
\end{figure}


Now consider adding particle $9$ to the residue $(3)(6)(7)(8)_{7}^{4}$ with an inverse soft factor in order to form the composite nine-particle residue $(3)(6)(7)(8)(9)_{79}^{4}$. Following the results of section~\ref{LeadingSingularities} this adds an additional particle to an existing MHV vertex in the channel diagram - see figure~\ref{fig:ISLcomposite678}. Adding particles $\{10,\ldots,n\}$ in the same way leads to the composite residue $(3)(6)\ldots(n)_{79\ldots n}^{4}$ which is the same function of momentum twistors as in equation~\eqref{standard3678} and whose channel diagram simply has more external legs on an MHV vertex. The residue is then independent of the momentum twistors $\{W_7,W_9,W_{10},\ldots,W_n\}$ in agreement with the collinear localisation. 


Let us now consider a second example of BCFW terms arising from a composite eight-particle residue. When minors $(1)$ and $(2)$ vanish we have the factorisation
\[
		(1)(2) \Rightarrow \left\{
		\begin{array}{c}
		(1)(2)\,^{6}\\
		(1)(2)\,_{2}\\
		\end{array} \right.	\, 	
\]

\begin{figure}[htp]
\centering
\includegraphics[height=6cm]{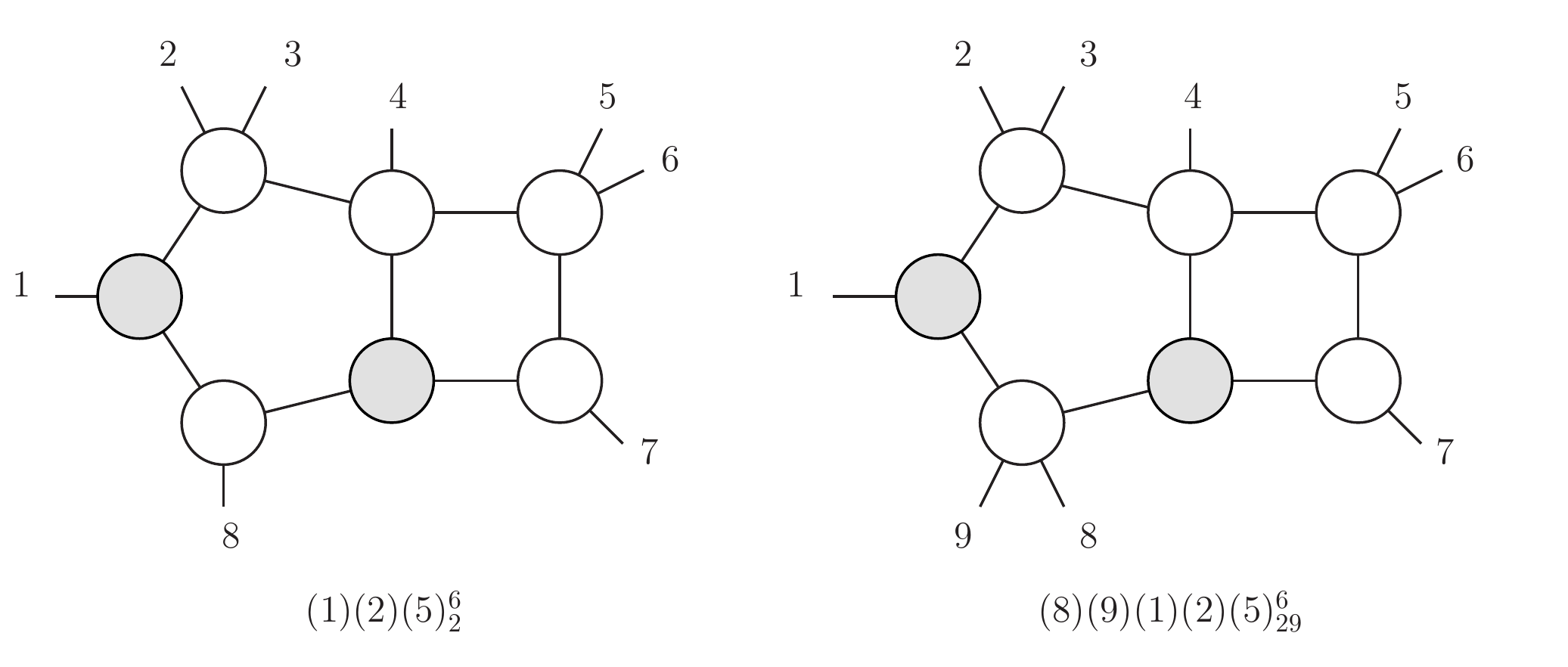}
\caption{\emph{Adding the particle $9$ to the residue  $(1)(2)(5)_2^6$ with an inverse soft factor.}}
\label{fig:ISLcomposite125}
\end{figure}

\ni and may impose the conditions that both $\{81234\}$ are coplanar and that $\{123\}$ are collinear. This is three conditions on two minors and therefore composite residues of the form $(1)(2)(i)_{2}^{6}$ may be defined. Here we consider the residue $(1)(2)(5)_{2}^{6}$ where the minor $(5)$ vanishes so that in addition $\{4567\}$ are coplanar - see figure~\ref{fig:ISLcomposite125}. Then we have the following expression in momentum twistors:
\begin{equation}
(1)(2)(5)_{2}^{6} = R(1,3,4,7,8)\,R(U,4,5,6,7) \qquad U = \la 1,3,4,[7\ra8]\, .
\end{equation}

\ni Adding particle 9 with inverse soft factors again adds an additional particle to one of the MHV vertices in the channel diagram and corresponds to the residue $(8)(9)(1)(2)(5)_{29}^{6}$ which is the same function of momentum twistors - see figure~\ref{fig:ISLcomposite125}. Again further particles $\{10,\ldots,n\}$ may be added in the same way.

\begin{figure}[htp]
\centering
\includegraphics[height=9cm]{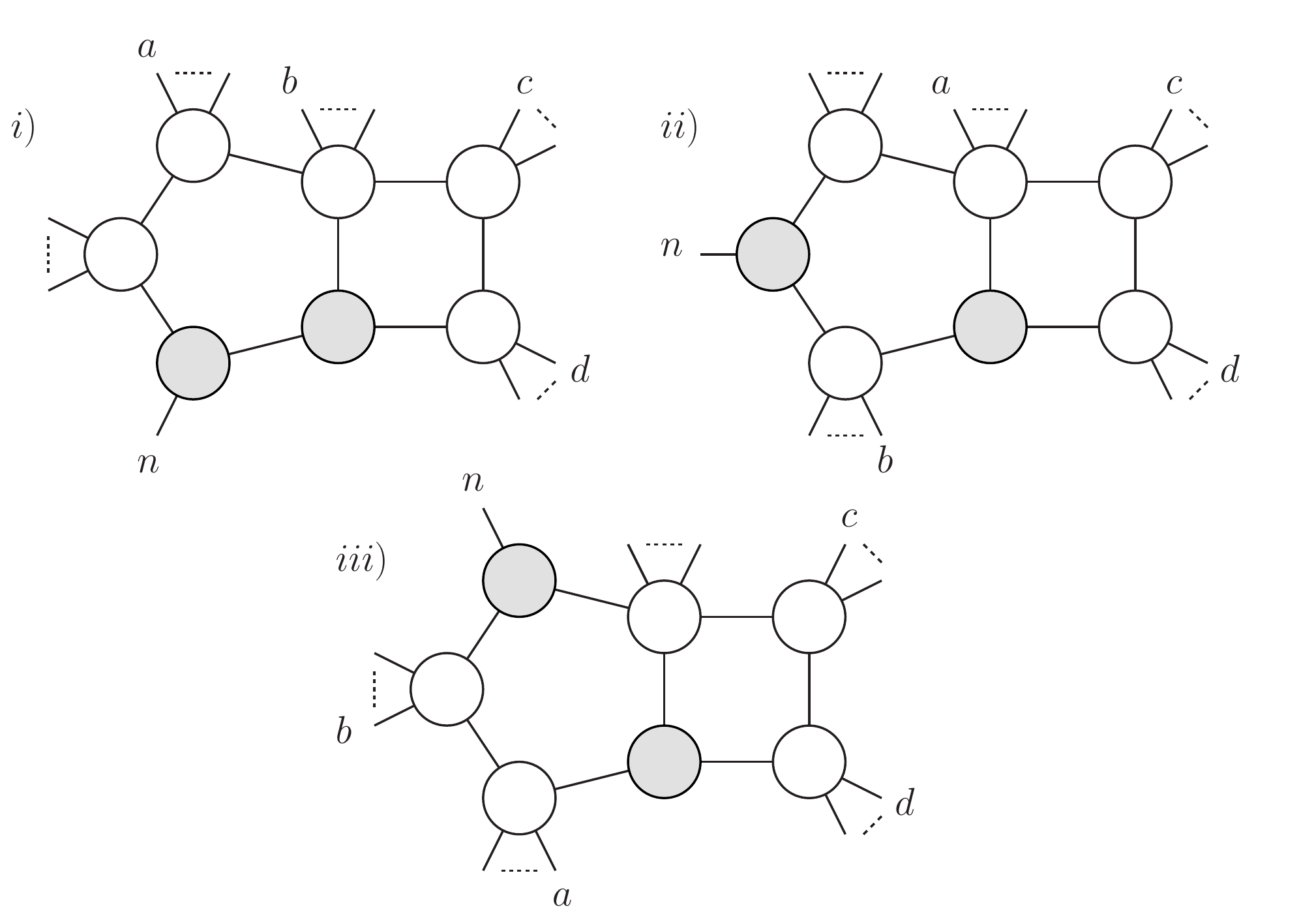}
\caption{\emph{The generic primitive pentabox channel diagrams appearing in BCFW expressions.}}
\label{fig:Pentaboxes1}
\end{figure}

The two above examples illustrate the construction of BCFW type residues with inverse soft factors. Starting with different eight-particle residues, all such BCFW terms can be found in this way and we can write down general expressions for the residues. The generic BCFW channel diagrams are those shown in figure~\ref{fig:Pentaboxes1} and their images under the reversal of particle labels $i\longrightarrow n-i$. Following~\cite{Drummond:2008cr} we have the notation

\begin{eqnarray}
		R_{n;ab;cd}  \equiv R(U,c-1,c,d-1,d) \nn\\	
		R_{n;ba;cd} \equiv R(V,c-1,c,d-1,d)
\end{eqnarray}

\ni for particular dual superconformal invariants where we have defined the momentum twistors
\begin{equation}
\label{UandV}
		 U \equiv \la n,a-1,a,[b-1 \ra W_b ] \hspace{0.5cm}\mbox{and}\hspace{0.5cm} V  \equiv \la n,b-1,b,[a-1 \ra W_a ] \, ,
\end{equation}

\ni which are naturally associated with on-shell loop momenta in the corresponding channel diagrams. Then the leading singularities in figure~\ref{fig:Pentaboxes1} correspond to the following residues
\begin{eqnarray}
		(i)&& R_{n;ab}R_{n;cd} = \overline{(a-1)(b-1)(c-1)(d-1)}_{\overline{\{ n,a-1,a,b-1,b,c-1,c,d-1,d\}}} \nn\\
		(ii)&& R_{n;ab}R_{n;ab;cd} = \quad \overline{(a-1)(c-1)(d-1)(b-1)}_{\overline{\{n,a-1,a,c-1,c,d-1,d,a-1,a\}}} \nn\\
		(iii)&& R_{n;ab}R_{n;ba;cd} = \quad \overline{(c-1)(d-1)(a-1)(b-1)}_{\overline{\{n,c-1,c,d-1,d,a-1,a,b-1,b\}}}
\end{eqnarray}

\ni where again overlines denote the complement of the enclosed set in $\{1,\ldots,n\}$.


\subsubsection{The Boundary BCFW Terms}
\label{BCFW Boundary}


We now consider so called boundary terms in BCFW solutions for the $\Nsq$ tree superamplitude~\cite{Drummond:2008cr}. These correspond to primitive leading singularities whose channel diagrams are pentaboxes where the shared MHV vertex has no external legs\cite{Bullimore:2009cb}. 


Let us now consider again eight-particle residues where the three adjacent minors $(6)$, $(7)$ and $(8)$ vanish and we have the factorisation of the two adjacent pairs,
\[
		(6)(7) \Rightarrow \left\{
		\begin{array}{c}
		(6)(7)^{3}\\
		(6)(7)_{7}\\
		\end{array} \right.		
\]
\[
		(7)(8) \Rightarrow \left\{
		\begin{array}{c}
		(7)(8)^{4}\\
		(7)(8)_{8}\\
		\end{array} \right.	\, .	
\]

\ni Now consider the composite residue $(6)(7)(8)_{78}^{34}$ where the two collinear subscript labels imply that the minor $(7)$ is vanishing to second order (see figure~\ref{fig:ISLcomposite678}). Note that the points $\{6781\}$ being collinear automatically implies that $\{56781\}$ and $\{67812\}$ are coplanar. In momentum twistors we have
\begin{equation}
(6)(7)(8)_{78}^{34} = R(1,2,3,5,6)\,R(U,V,3,4,5)
\end{equation}

  \ni where have defined momentum twistors $U=\la 1,2,3,[5\ra 6]$ and $V=\la1,5,6,[2\ra3]$ again associated with fixed on-shell loop momenta. Note that even for such boundary terms the result may be written simply as a product of two basic R-invariants; a result that is made transparent by performing the generalised unitarity calculation directly in momentum twistor space~\cite{forthcoming}. It is then immediate to add particles $\{9,\ldots,n\}$ to an MHV vertex in the channel diagram corresponding to the residue $(6)(7)\ldots(n)_{78\ldots n}^{34}$ which is the same function of momentum twistors.
 
\begin{figure}[htp]
\centering
\includegraphics[height=6cm]{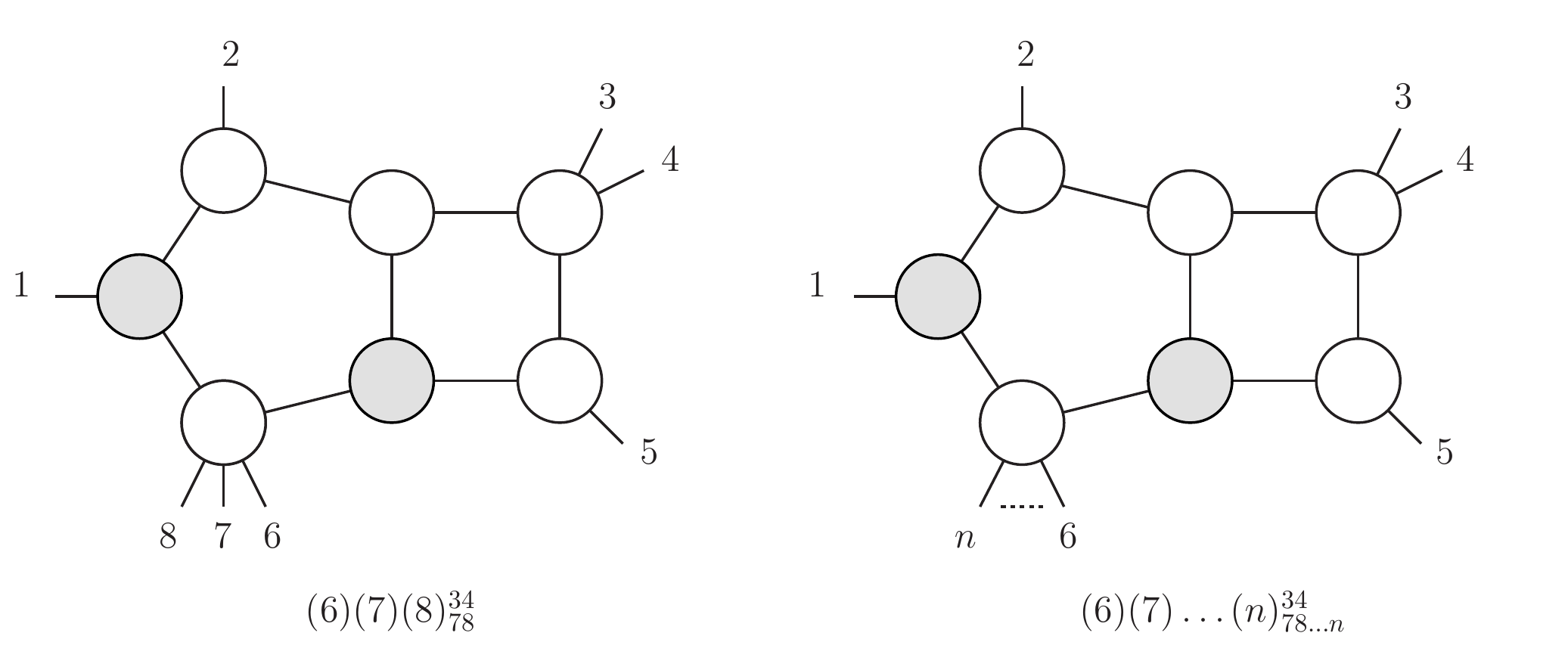}
\caption{\emph{An example of inverse soft factors applied to a BCFW boundary term.}}
\label{fig:ISLstandard7812}
\end{figure}
 
 
We now consider a second example of BCFW boundary terms that are standard residues of the eight-particle grassmannian formula of the form $(7)(8)(1)(2)$. The three pairs of adjacent minors factorise as follows

\[
		(7)(8) \Rightarrow \left\{
		\begin{array}{c}
		(7)(8)\,^{4}\\
		(7)(8)\,_{8}\\
		\end{array} \right.		
\]
\[
		(8)(1) \Rightarrow \left\{
		\begin{array}{c}
		(8)(1)\,^{5}\\
		(8)(1)\,_{1}\\
		\end{array} \right.		
\]
\[
		(1)(2) \Rightarrow \left\{
		\begin{array}{c}
		(1)(2)\,^{6}\\
		(1)(2)\,_{2}\\
		\end{array} \right.		
\]

\ni and there are two solutions not imposing singular kinematics, which are the residues $(7)(8)(1)(2)_{82}^{5}$ and $(7)(8)(1)(2)_{1}^{46}$ (see figure~\ref{fig:ISLstandard7812}). The momentum twistor expressions are, for example,
\begin{eqnarray}
(7)(8)(1)(2)_{1}^{46} = R(2,3,4,5,6)\, R(U,V,6,7,8)
\end{eqnarray}

\ni where we have defined $U = \la 4,5,6,[2\ra3]$ and $V=\la2,3,4,[5\ra6]$. Again adding further particles $\{9,\ldots,n\}$ to an MHV vertex leaves the dual superconformal invariant unchanged as a function of momentum twistors.

\begin{figure}[htp]
\centering
\includegraphics[height=6cm]{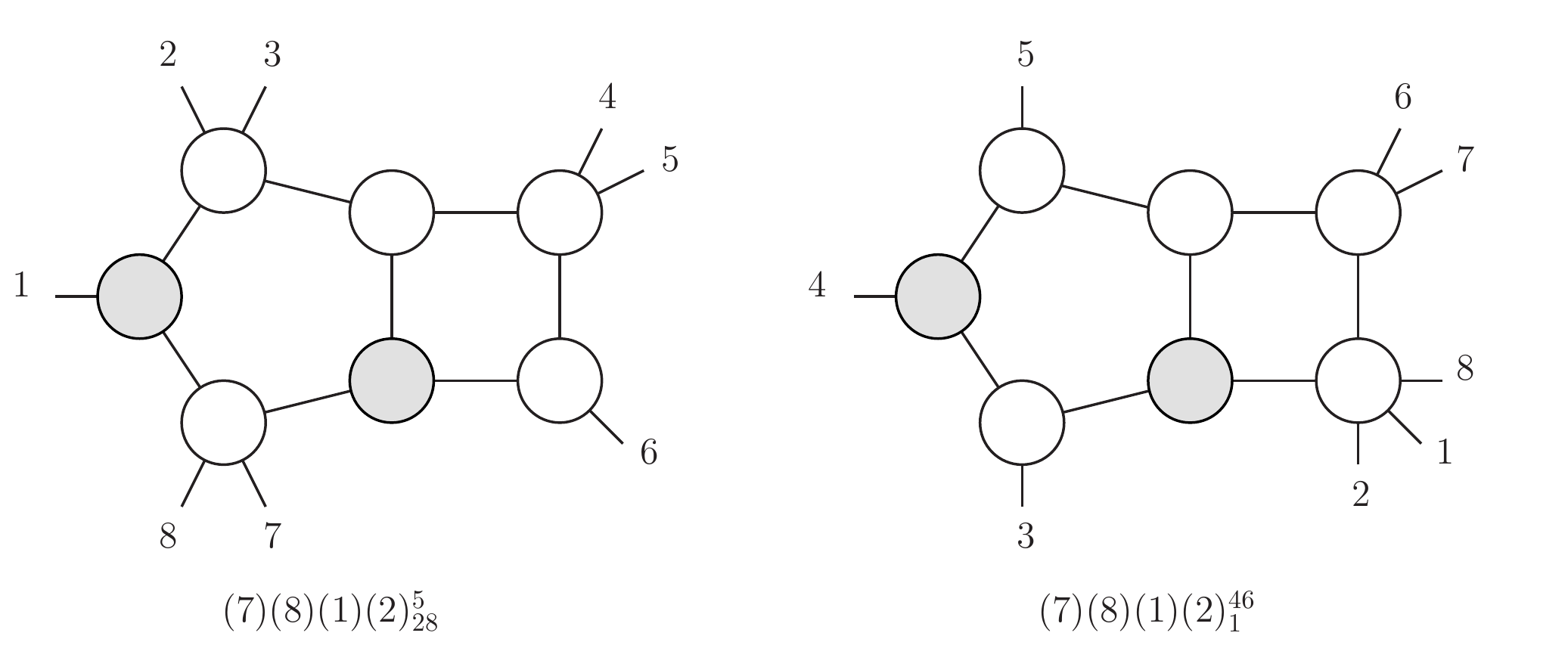}
\caption{\emph{Further examples of BCFW boundary terms and the corresponding residues.}}
\label{fig:ISLstandard7812}
\end{figure}

\begin{figure}[htp]
\centering
\includegraphics[height=5cm]{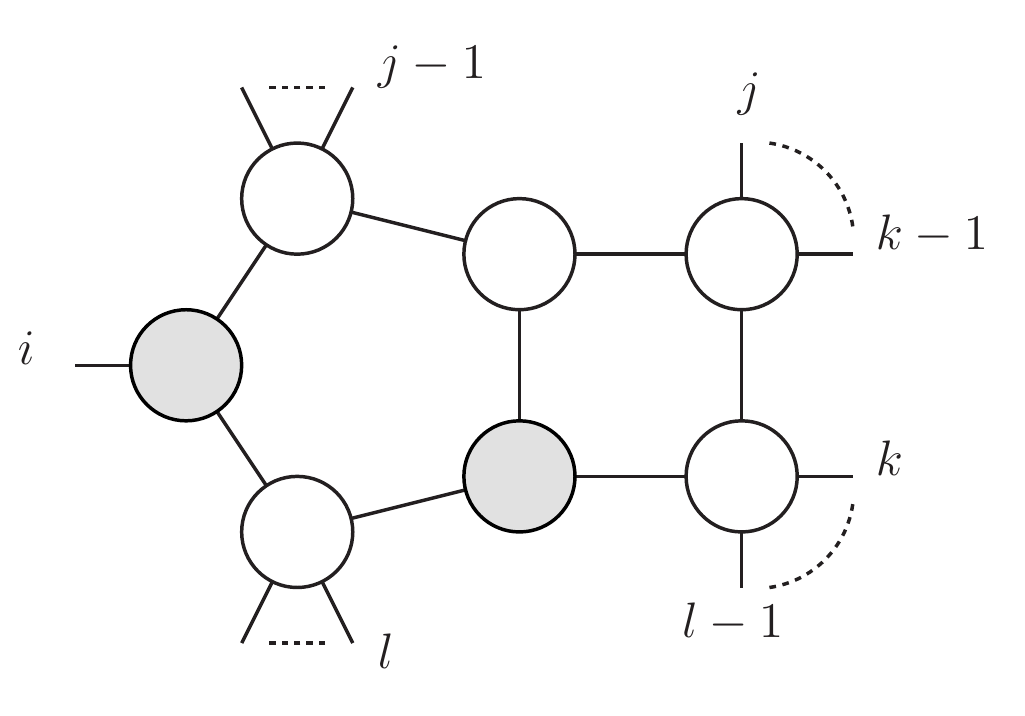}
\caption{\emph{The generic BCFW boundary term found from the residue $(7)(8)(1)(2)_{1}^{46}$ by applying multiple inverse soft factors.}}
\label{fig:Standard7812gen}
\end{figure}

Starting from the eight-particle residue $(7)(8)(1)(2)_{1}^{46}$ we can construct the generic BCFW boundary term with the same channel diagram by inverse soft factors - see figure~\ref{fig:Standard7812gen}. The momentum twistor expression is always a product of two basic R-invariants with shifted arguments and in this case we have

\begin{equation}
\overline{(j-1)}_{\overline{\{i,j-1,j,k-1,k,l-1,l\}}} = R(i,j-1,j,l-1,l)\,R(U,V,j,k-1,k)
\end{equation}

\ni where we have defined the following momentum twistors 
\begin{equation}
U=\la i,j-1,j,[l-1\ra W_l] \qquad \mathrm{and} \qquad V=\la i,l-1,l,[j-1\ra W_j].
\end{equation}

\ni that are associated to cut propagators in the channel diagram. Similarly one can construct by inverse soft factors all such boundary terms corresponding to degenerate cases of the pentaboxes $(i)$,$(ii)$ and $(iii)$ in figure~\ref{fig:Pentaboxes1} from the previous subsection.

\subsubsection{Kissing Boxes}
\label{Kissingboxes}

Here we will consider residues corresponding to kissing box channel diagrams that appear as leading singularities of two-loop amplitudes. Consider standard eight-particle residues of the form $(2)(3)(5)(6)$, then each pair of minors factorises as follows, 

\[
		(2)(3) \Rightarrow \left\{
		\begin{array}{c}
		(2)(3)\,^{7}\\
		(2)(3)\,_{3}\\
		\end{array} \right.		
\]
\[
		(5)(6) \Rightarrow \left\{
		\begin{array}{c}
		(5)(6)\,^{2}\\
		(5)(6)\,_{6}\\
		\end{array} \right.		
\]\vspace{0.1mm}

\begin{figure}[htp]
\centering
\includegraphics[height=5.5cm]{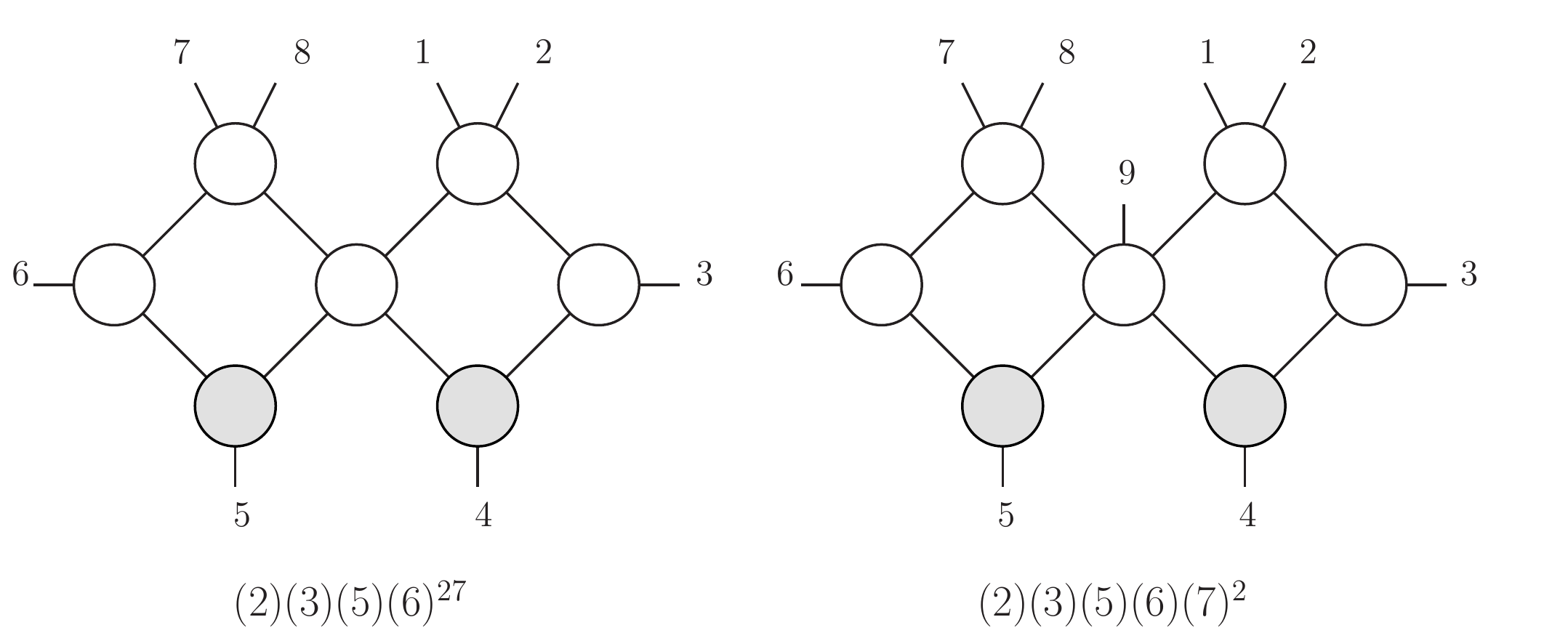}
\caption{\emph{Examples of residues with kissing box channel diagrams that may not be constructed by inverse soft factors from lower point leading singularities.}}
\label{fig:Standard2356}
\end{figure}

\ni and there are here two solutions $(2)(3)(5)(6)^{27}$ and $(2)(3)(5)(6)_{36}$ not imposing singular kinematics. The first residue has localisation where the points $\{45678\}$ and $\{12345\}$ are collinear and corresponds to a kissing box channel diagram shown in figure~\ref{fig:Standard2356}. Here the residue may easily be calculated from generalised unitarity with the result,
\begin{equation}
		(2)(3)(6)(7)^{27} = R(5,6,7,8,1)\,R(8,1,2,3,4).
\end{equation}

First we take the opportunity to study a residue that cannot be constructed by inverse soft factors. Consider then the nine-particle residue $(2)(3)(5)(6)(7)^{2}$ whose channel diagram is shown in figure~\ref{fig:Standard2356}. It is clear from the channel diagram that any attempt to construct this residue from $(2)(3)(6)(7)^{27}$ by adding particle 9 will result in a three-loop leading singularity. The same conclusion is also clear from the absence of a collinearity subscript. Such new channel diagrams, not constructible by inverse soft factors, appear with each additional particle.


\begin{figure}[htp]
\centering
\includegraphics[height=5.5cm]{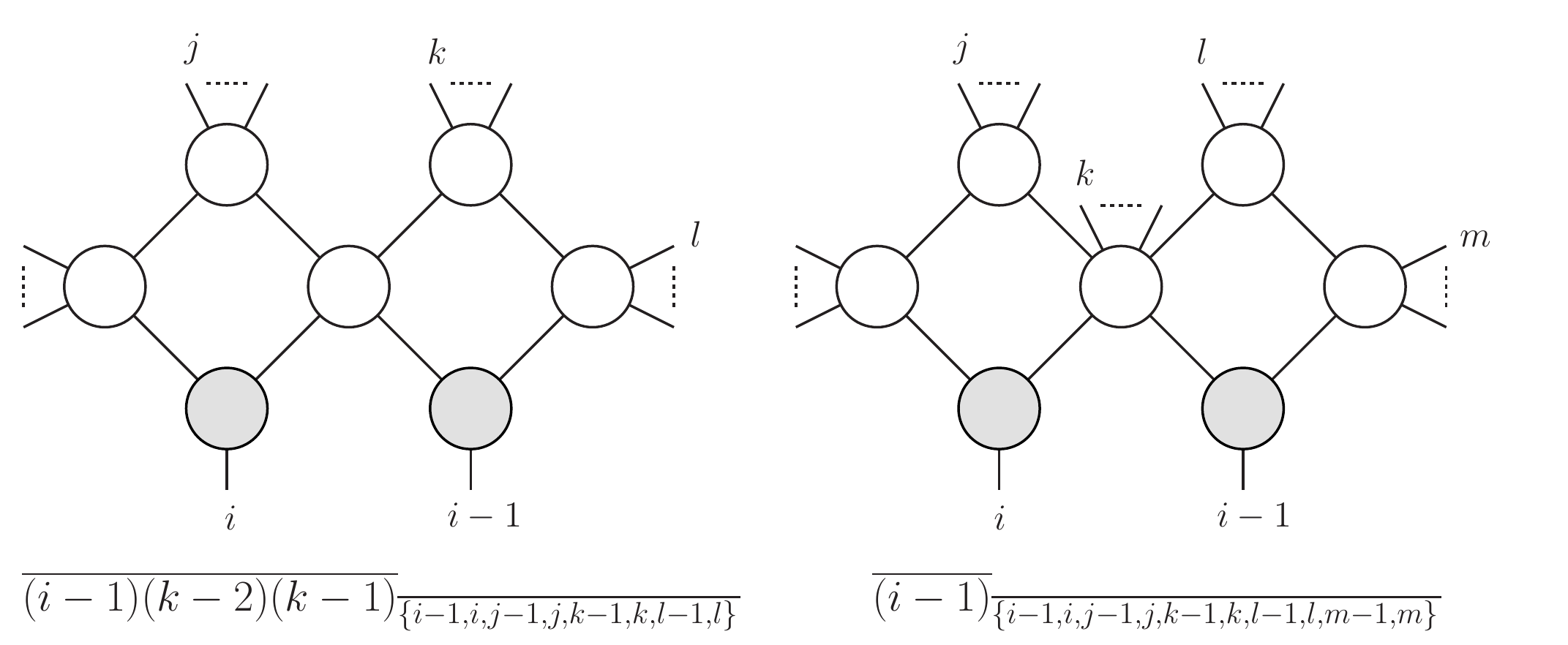}
\caption{\emph{Examples of generic kissing box channel diagrams and their residues. }}
\label{fig:Standard2356gen}
\end{figure}

The generic kissing box channel diagrams may all be constructed by inverse soft factors, but generically the process must be start from a residue with more than eight particles. The most generic kissing boxes constructed from the residues $(2)(3)(6)(7)^{27} $ and $(2)(3)(5)(6)(7)^{2}$ by inverse soft factors are shown in figure~\ref{fig:Standard2356gen}.


\subsubsection{Four-mass Box Coefficients}

\begin{figure}[htp]
\centering
\includegraphics[height=5.5cm]{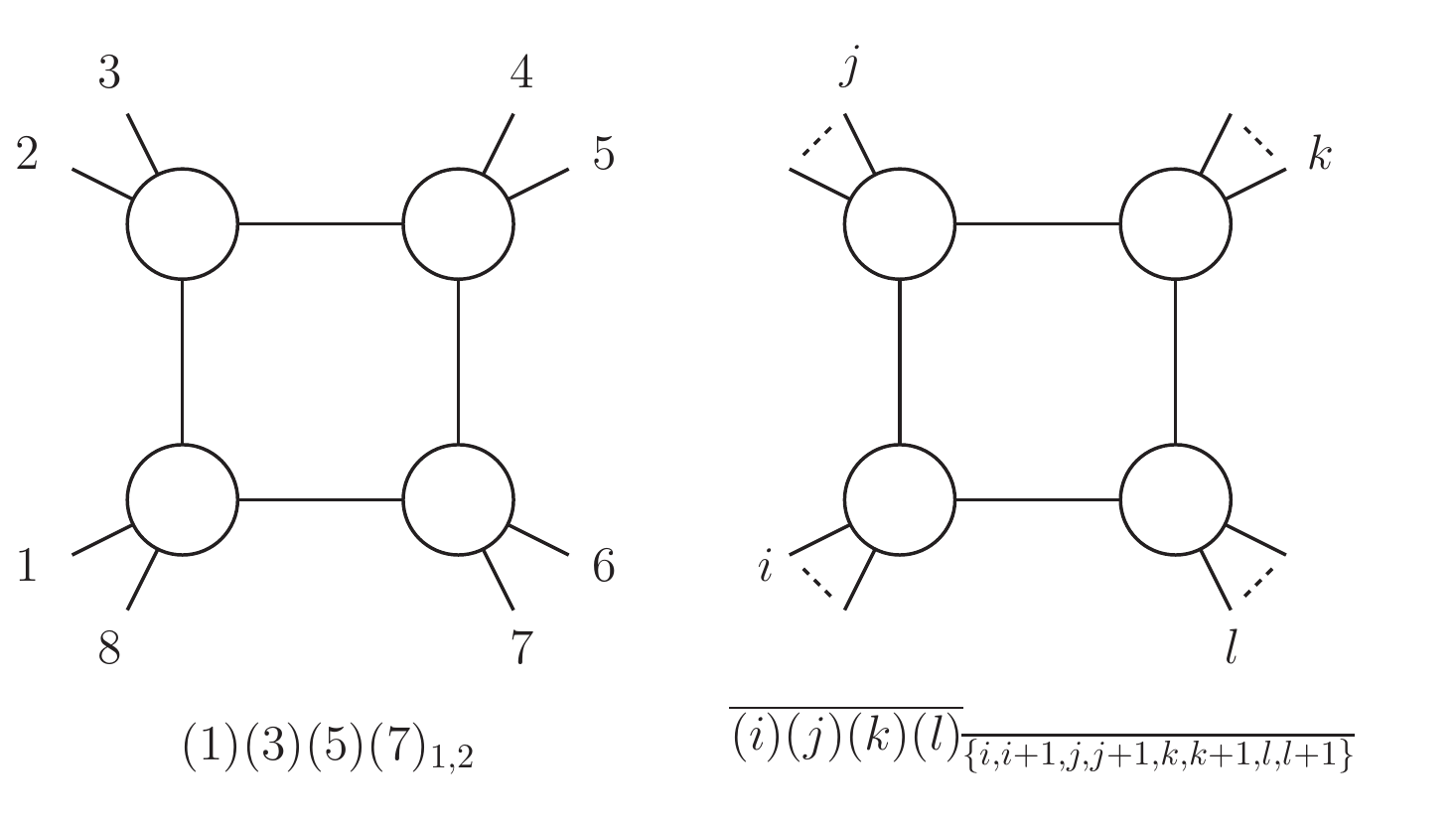}
\caption{\emph{The generic four-mass box coefficients and the corresponding grassmannian residues.}}
\label{fig:Fourmassbox}
\end{figure}

Finally we consider residues corresponding to four-mass box coefficients - see figure~\ref{fig:Fourmassbox}. We start from the eight-particle residues $(1)(3)(5)(7)_{1,2}$ where the subscript indicates one of two solutions to the cut conditions for the four mass box configuration~\cite{ArkaniHamed:2009dn}. These residues have the following expressions in terms of momentum twistors~\cite{Mason:2009qx},
\begin{equation}
R(X_{(i)},3,4,7,8)\,R(Y_{(i)},1,2,5,6).
\end{equation}

\ni The momentum twistors $X_{(i)}$ and $Y_{(i)}$ with $i=1,2$ are determined by the two solutions to the cut conditions of the four-mass box configuration and may be found in~\cite{Mason:2009qx}. Note however that there are many such representations of the four-mass box coefficients.

Since each MHV vertex in the channel diagram has two external states, particles may be added to all them forming generic four-mass box coefficients - see figure~\ref{fig:Fourmassbox}. From the general rule for inverse soft limits, the momentum twistor expressions are then

\begin{equation}
R(X_{(i)},j,j+1,l,l+1)R(Y_{(i)},i,i+1,k,k+1)\, ,
\end{equation}

\ni where again the momentum twistors $X_{(i)}$ and $Y_{(i)}$ are determined by the two solutions to the cut conditions. They may be obtained from the solutions for eight particles by replacing momentum twistors $\{1,\ldots,8\}$ with $\{i,i+1,j,j+1,k,k+1,l,l+1\}$ in accordance with the general rule for inverse soft factors. Adding particles in between legs on separated MHV vertices will of course lead to higher loop leading singularities.


\subsection{Higher Loops}

So far we have considered inverse soft factors that do not change the number of loops in the channel diagram. Here we will consider an example studied above and find the most generic residues that can be constructed by inverse soft factors. Hopefully this will serve as representative of a similar procedure for all residues.

\begin{figure}[htp]
\centering
\includegraphics[height=5.5cm]{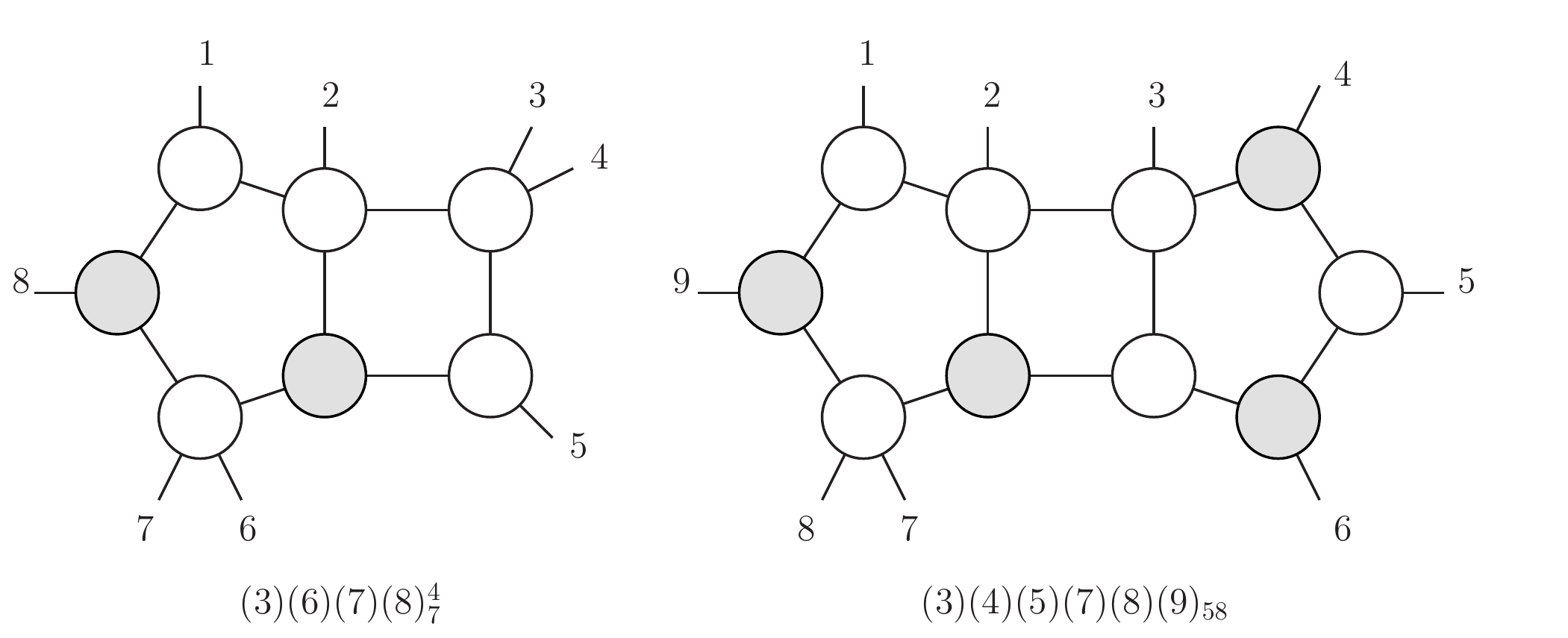}
\caption{\emph{An inverse soft factor that increases the number of loops in the channel diagram.}}
\label{fig:678threeloop}
\end{figure}

We consider the residue $(3)(6)(7)(8)_{7}^{4}$ studied in section~\ref{BCFWstandard} which corresponds to a standard BCFW type channel diagram. We first perform an inverse soft factor in between particles 4 and 5 which lie on separate MHV vertices in the channel diagram. After relabelling the external particles we find the following residue,
\begin{equation}
(3)(4)(5)(7)(8)(9)_{58} = R(9,1,2,6,7)\,R(U,2,3,4,5) \qquad U = \la 9,1,2,[6\ra7]\, ,
\end{equation}
which corresponds to a three-loop primitive channel diagram (see figure~\ref{fig:678threeloop}).

\ni Adding further particles to MHV vertices by inverse soft factors we find the following residue (see figure~\ref{fig:678threeloop})

\begin{equation}
\overline{(j-i)(j)(l)}_{\overline{\{i,j-1,j,j+1,k,l,l+1\}}} = R(i,j-1,j,l,l+1)\,R(U,j,j+1,k,l)\, 
\end{equation}

\begin{figure}[htp]
\centering
\includegraphics[height=7cm]{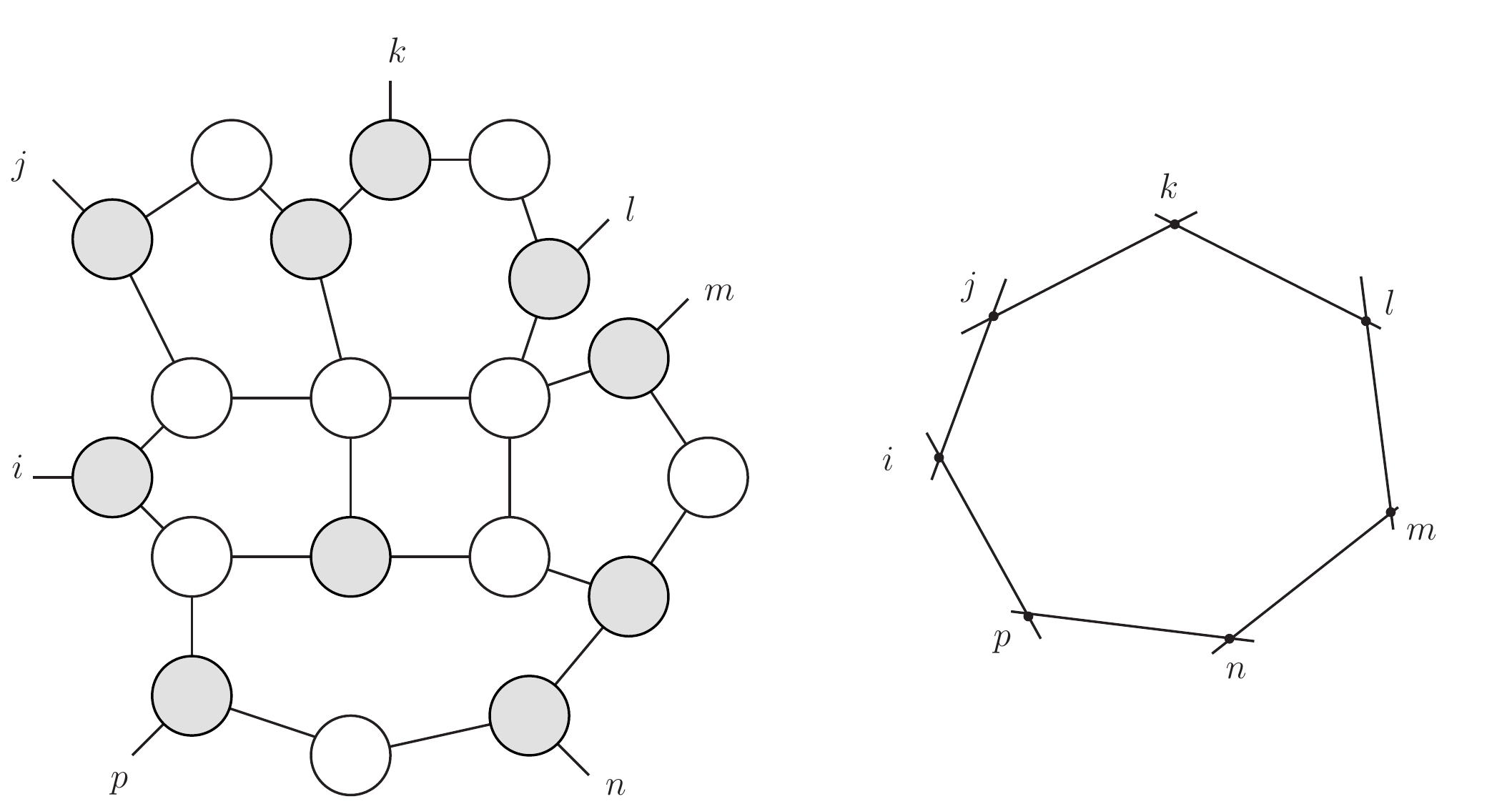}
\caption{\emph{The most generic channel diagram constructed from the residue $(3)(6)(7)(8)_{7}^{4}$ by inverse soft factors and it twistor space support.}}
\label{fig:678generic}
\end{figure}

\ni where we have defined the momentum twistor $U = \la i,j-1,j,l\ra l+1]$. Finally we can add further particles with inverse soft factors in between the remaining adjacent MHV vertices to form the most generic residue constructed by inverse soft factors (see figure~\ref{fig:678generic})
\begin{equation}
\overline{\{0\}}_{\overline{\{i,j-1,j,k,l,m,n,p\}}} = R(i,j,k,m,p)\,R(U,k,l,m,n)
\end{equation}

\ni where we now define $U = \la i,j,k,[n\ra p]$ and the notation $\overline{\{0\}}$ means that all minors vanish on this residue. All further inverse soft factors will simply add particles to existing MHV vertices without increasing the number of loops in the channel diagram.

For large numbers of particles, it is the generic situation that all of the minors are vanishing. The twistor support then consists of a series of $i\leq9$ lines containing generic numbers of particles, that intersect only with their neighbours in the chain. In such cases the channel diagram is highly non-unique and the residues may be reached in many ways by inverse soft limits. Any attempt to add further loops with inverse soft factors will now just add further particles to existing vertices and therefore channel diagrams with more than six loops do not lead to new Yangian invariants at $\Nsq$. More generally, similar arguments for N$^p$MHV amplitudes indicate that no new Yangian invariants appear for channel diagrams with more than $3p$ loops~\cite{Bullimore:2009cb}.


\section{Comments on Higher MHV Degree}

For N$^{k-2}$MHV amplitudes with $k\geq5$ the passage from grassmannian localisation in $\CP^{k-1}$ to twistor support is not as immediate. However, the vanishing of minors to various orders always leads directly to conditions on the grassmannian localisation in $\CP^{k-1}$. Therefore grassmannian localisation seems to be more fundamental in the description of residues. 

Consider N$^3$MHV scattering amplitudes where the vanishing of the minor $(2)=(12345)$ implies that the the points $\{12345\}$ lie in a 3-plane in $\CP^4$. Consider now that both minors $(2)$ and $(3)$ vanish so that both sets of points $\{12345\}$ and $\{23456\}$ lie in 3-planes in $\CP^4$. Then we have the following factorisation of minors corresponding to two ways this can happen:

\begin{enumerate}
\item The points $\{123456\}$ all lie in the same 3-plane in $\bP^4$
\item The points $\{2345\}$ lie in a 2-plane in $\bP^4$.
\end{enumerate}

\ni When the next adjacent minor $(4)$ vanishes, for example, we can place four conditions on three minors with the points $\{23456\}$ coplanar in $\CP^4$ and the minor $(3)$ vanishing to second order. However, for N$^3$MHV amplitudes the minors may vanish to third order. For example the minor $(1)$ can vanish to third order when points $\{234\}$ become collinear in $\CP^4$. Therefore to define residues at N$^3$MHV requires the specification of the vanishing minors that denote the sets of points lying in 3-planes, the sets of points lying in 2-planes, and the sets of collinear points in $\CP^4$. Such information should specify the grassmannian localisation and uniquely determine the residue at  N$^3$MHV.

More generally for N$^{k-2}$MHV amplitudes we expect that individual residues are completely determined by specifying the following grassmannian localisation properties in $\CP^{k-1}$:

\begin{itemize}
\item A list vanishing minors specifying sets of $k$ points that lie in $(k-2)$-planes
\item The sets of $(k-1)$ points that lie in $(k-3)$ planes
\item $\ldots$
\item The sets of four points that are coplanar
\item The sets of three points that are collinear.
 \end{itemize}

Following results presented here at $\Nsq$ further examples at N$^3$MHV, it is natural to conjecture that all N$^{k-2}$HMV residues may be written as a product of $(k-2)$ basic dual superconformal invariants $R(U,V,X,Y,Z)$ for some choice of arguments that are associated with cut propagators in the corresponding channel diagrams. We leave for further investigation the action of inverse soft limits on residues of higher MHV degree.

\section*{Acknowledgements} It is a pleasure to thank Andrew Hodges, Lionel Mason and David Skinner for many useful discussions and for their encouragement and advice. I would also like to thank the Institute for Advanced Study in Princeton and in particular Nima Arkani-Hamed for their hospitality in April when part of this work was carried out. I am supported by an STFC studentship.

\bibliographystyle{JHEP-2}
\bibliography{Amplitudes}

\end{document}